
\raggedbottom


\def\nex{\par\noindent\hang}

\null
\vskip 20mm
\centerline{\bf Star Cluster Evolution,
Dynamical Age Estimation}

\vskip 2mm

\centerline{\bf and the
Kinematical Signature of Star Formation}

\vskip 2mm

\centerline{\bf K3}

\vskip 10mm
\centerline{\bf Pavel Kroupa}
\vskip 10mm
\centerline{Astronomisches Rechen-Institut}
\vskip 2mm
\centerline{M{\"o}nchhofstra{\ss}e~12-14, D-69120~Heidelberg, Germany}
\vskip 10mm
\centerline{e-mail: S48@ix.urz.uni-heidelberg.de}

\vfill

\centerline{MNRAS, in press}

\vfill\eject

\hang{
\bf Abstract.
\rm
We distribute 400 stars in $N_{\rm bin}=200$ binary systems in clusters
with initial half mass radii $0.077\le R_{0.5}\le 2.53\,$pc and follow the
subsequent evolution of the stellar systems by direct N-body integration. The
stellar masses are initially paired at random from the KTG(1.3)
initial stellar mass function. The initial period distribution is
flat ranging from $10^3$ to $10^{7.5}$~days, but we also perform simulations
with a
realistic distribution of periods which rises with increasing $P>3$~days and
which is consistent with pre-main sequence observational constraints.
For comparison we simulate the evolution
of single star clusters.
After an initial relaxation phase,
all
clusters evolve according to the same $n(t)\propto {\rm exp}(-t/\tau_{\rm e})$
curve, where
$n(t)$ is the number density of stars in the central 2~pc sphere at time
$t$ and $\tau_{\rm e}\approx230$~Myrs. All clusters
have the same lifetime $\tau$. n(t) and $\tau$ are thus independent
of (i) the inital proportion of binaries and (ii) the initial $R_{0.5}$. Mass
segregation measures the dynamical age of the cluster: the mean
stellar mass inside the central region increases approximately linearly with
age. The
proportion of binaries in the central cluster region is a sensitive indicator
of the initial cluster concentration: it declines within approximately the
first 10--20 initial relaxation times
and rises only slowly with age, but for initial $R_{0.5}<0.8$~pc, it is
always significantly
larger than the binary proportion outside the central region.
If most stars form in binaries in embedded clusters that are dynamically
equivalent
to a cluster specified initially by $(N_{\rm bin},R_{0.5})=(200,0.85\,{\rm
pc})$, which is located at the edge of a $1.5\times10^5\,M_\odot$
molecular cloud with a diameter of 40~pc,
then we estimate that at most about 10~per cent of all pre-main
sequence stars achieve near escape velocities from the molecular cloud.
The large ejection velocities resulting from close
encounters between binary
systems imply a distribution of young stars over
large areas surrounding star forming sites.
This `halo' population of a molecular cloud complex
is expected to have a significantly reduced binary proportion (about 15~per
cent or less) and a significantly increased proportion of stars with depleted
or completely removed circumstellar disks. In this scenario, the distributed
population is expected to have a similar
proportion of binaries as the Galactic field (about 50~per cent).
If a distributed population shows orbital parameter
distributions not affected by stimulated evolution (e.g. as in Taurus--Auriga)
then it probably originated in a star-formation mode in which the binaries form
in relative isolation rather than in embedded clusters.
The Hyades Cluster luminosity function suggests an advanced dynamical age. The
Pleiades luminosity function data suggest a distance modulus $m-M=6$, rather
than 5.5.
The total proportion of binaries in the central region of the Hyades and
Pleiades Clusters are probably 0.6--0.7.
Any observational luminosity function of a Galactic cluster must be corrected
for unresolved binaries when studying the stellar mass function.
Applying our parametrisation for open
cluster evolution we estimate the birth masses of both clusters.
We find no evidence for
different dynamical properties of stellar systems at birth in the Hyades,
Pleiades and Galactic field stellar samples.
Parametrising the depletion of low
mass stars in the central cluster region by the ratio, $\zeta(t)$, of the
stellar luminosity
function at the `H$_2$--convection peak' ($M_{\rm V}\approx12$) and
`H$^-$ plateau' ($M_{\rm V}\approx7$),
we find good agreement with the Pleiades and Hyades $\zeta(t)$
values. The observed proportion of binary stars in the very young Trapezium
Cluster is consistent with the early dynamical evolution of a cluster with a
very high initial stellar number density.

\vskip 5mm

}

\vfill\eject

\noindent{\bf 1 INTRODUCTION}
\vskip 12pt
\noindent
We refer to the stellar mass function, the proportion of stellar
systems (singles, binaries, triples, etc.) and the distribution of their
orbital parameters as the {\it dynamical properties of stellar systems}.
Studying Galactic clusters is important because they represent fossils of
discrete star formation events. From them we can hope to obtain information on
the dependence of the dynamical properties of stellar systems on the birth
conditions.

The majority of stars in the Galactic disk may be born in embedded clusters
rather than in isolation. This conclusion is drawn by Lada \& Lada (1991)
after studying the distribution of young stellar objects in the
L1630 molecular cloud of the Orion complex. In it they find no
significant distributed population of young stars but four embedded clusters
containing at least about 627 objects. Strom, Strom \&
Merrill (1993), on the other hand, find a distributed population of about 1500
young stars in the L1641 molecular cloud in Orion, as well as seven clusters,
in which 10--50 young stellar objects have been detected, and one partially
embedded cluster, in which 150 young stellar objects have been detected.
The ages ($<1$~Myrs) of the stellar objects in the detected clusters
appear to be younger than in the distributed population
(5--7~Myrs). This is consistent with the latter having originated in
aggregates which are now dissolved. Nevertheless, it is
evident that the observational evidence for a predominant clustered star
formation mode remains suggestive rather than conclusive.
However, if it is assumed that star formation nearly always produces a binary
star (as suggested by observations of pre-main sequence stars)
then clustered star formation {\it must} be the dominant
mode, rather than distributed star formation (Kroupa 1995a, hereafter K1).

This paper is the third in a series of three papers K1, K2 and K3. In K1 we
study the evolution of a binary star population in stellar aggregates and use
the results for inverse dynamical population synthesis on the observed
dynamical properties of stellar systems in the Galactic field to deduce that
a dominant clustered mode of star formation may exist. In K1 we suggest that
most Galactic field stars may have been born in aggregates that are dynamically
equivalent to the `dominant mode cluster', which is defined by $(N_{\rm
bin},R_{0.5})=(200,0.85\,{\rm pc})$, where $N_{\rm bin}$ is the
initial number of binaries and $R_{0.5}$ is the initial half mass radius. We
also derive
an initial distribution of periods. In K2 (Kroupa 1995b) we study in detail the
dynamical properties of Galactic field stellar systems if they form
in the dominant mode cluster. In this paper (K3) we make use of the N-body
simulations of K1 and K2 to study the overall evolution of the
stellar clusters.

The realistic KTG(1.3) stellar mass function (Section~2) is adopted,
and all stars are initially in aggregates, or clusters, of binary systems.
These have a period and mass ratio
distribution consistent with pre-main sequence data. Our initial very large
proportion of primordial binaries is also consistent with observational
evidence that a large proportion of binaries may reside in the central region
of at least one Galactic cluster (the Praesepe Cluster, Kroupa
\& Tout 1992). As a control experiment we
simulate single-star clusters. We are interested
to learn if and how the initial conditions of our simulations are observable in
real Galactic clusters. By studying the rate at which the clusters dissolve
and the evolution of mass and binary star segregation the evolutionary
history of a Galactic cluster can probably be traced. The stellar
luminosity function in the central regions
of clusters evolves as a result of mass segregation, evaporation of
stars and increase of the proportion of binary systems.
We also investigate if the distribution of stellar systems in space and in
velocity after
dissolution of the aggregates can be used to obtain clues about the
initial dynamical configuration.

We concentrate on readily observable properties of stellar
clusters and
do not mean to provide a comprehensive treatment of their evolution
which has been extensively studied elsewhere:
Mathieu (1985) describes the internal kinematics and the structure of Galactic
clusters, Wielen (1985) discusses their dynamics, and
Aarseth (1988a,b) reviews the dynamical evolution of open clusters and
discusses core collapse, respectively, based on the extensive and detailed
simulations of Terlevich (1987).
The ejection of stars from open
clusters containing binaries is studied in the context of OB runnaway
stars by Leonard \& Duncan (1990). The hypothesis that
blue stragglers may result from collisions and merging of finite sized stars in
clusters is discussed by Leonard \& Linnell (1992).
Hut (1985) emphasises that binaries can be considered a dynamical energy source
in a cluster similar to nuclear reactions in a star.
A comprehensive review of the role of
binaries in Globular cluster dynamics is provided by Hut et al. (1992).
Heggie \& Aarseth (1992) and McMillan \& Hut (1994) consider the
evolution of globular clusters consisting of equal-mass stars and containing
an initial proportion of binaries of up to 20~per cent. These authors, and the
model of the binary star evolution in a globular cluster devised by Hut,
McMillan \& Romani (1992), provide many
valuable insights into the processes that govern the interaction of the cluster
with the population of binary systems.

In Section~2 we briefly summarise the assumptions, definitions and our method.
The evolution of four binary star clusters and two single star clusters is
discussed in Section~3 and compared with the evolution of the dominant mode
cluster in Section~4, where we also elaborate on the concept of `dynamical
equivalence' introduced in K1. The kinematical signature of star formation is
addressed
in Section~5. In Section~6 we apply our simulations to three clusters and
discuss the stellar luminosity function. We conclude with Section~7.

\nobreak\vskip 10pt\nobreak
\noindent{\bf 2. ASSUMPTIONS, DEFINITIONS AND METHOD}
\nobreak\vskip 10pt\nobreak
\noindent
A detailed description of our assumptions and method can be found in section~3
of K1, which we briefly summarise here.

We distribute $N_{\rm bin}=200$ binary systems according to a Plummer density
law with half mass radii $R_{0.5}=0.077, 0.25, 0.77, 2.53$~pc.
$R_{0.5}=0.08$~pc
corresponds to an initially highly concentrated cluster similar to the
Trapezium Cluster, whereas $R_{0.5}=2.53$~pc corresponds to a loose cluster
which approximates distributed star formation (e.g. Taurus--Ariga). We also
distribute $N_{\rm sing}=400$ single stars in clusters with $R_{0.5}=0.077,
0.25$~pc. These are of academic interest only and serve as a comparison with
the realistic binary star clusters. Virial equilibrium is assumed.
For the N-body simulation of the dynamical evolution
of each cluster we use
the program {\it nbody5} written by Aarseth (1994).
We model a standard Galactic tidal field (see K1).

Stellar masses in the range $0.1\,M_\odot\le m\le 1.1\,M_\odot$ are obtained
from the initial mass function which is
conveniently approximated by $\xi(m)\propto m^{-\alpha_i}$,
$\alpha_1=1.3$ for $0.08\,M_\odot \le m<0.5\,M_\odot$, $\alpha_2=2.2$ for
$0.5\,M_\odot \le m<1.0\,M_\odot$, $\alpha_3=2.7$ for $1.0\,M_\odot\le m$
and $\xi(m)\,dm$ is the number of stars with masses in the range $m$ to $m+dm$
(Kroupa et al. 1993). We refer to $\xi(m)\,dm$ as the KTG($\alpha_1$) mass
function. Our adopted stellar mass range allows
us to ignore post-main sequence stellar evolution which simplifies the
computation and allows focusing on purely stellar dynamical evolution.
If we have a population of stars with masses $0.1\,M_\odot
\le m < {\cal A}$, then for ${\cal A}=10\,(50)\,M_\odot$, about 8~per cent of
these are more massive than
$1.1\,M_\odot$, and contribute 35~(39)~per cent to the total mass.
Stars more massive than about $5~(10)\,M_\odot$ will affect the dynamical
evolution of the clusters within the first $7\times10^7~(2\times10^7)$~yrs, but
will be insignificant during most of the lifetime of the clusters studied here.
The mean
stellar mass in our models is $0.32\,M_\odot$ and the mass of each cluster is
$128\,M_\odot$, which is near the peak of the mass function of Galactic
clusters
(Battinelli, Brandimarti \& Capuzzo-Dolcetta 1994).

To build the initial binary stars
we combine the stellar masses at random and distribute orbital periods, $P$ (in
days), from a flat distribution, $f_{\rm P}({\rm log}_{10}P)=\left[{\rm
log}_{10}(P_{\rm max}) - {\rm log}_{10}(P_{\rm min})\right]^{-1}$
(equation~3 in K1), with log$_{10}P_{\rm min}=3$, log$_{10}P_{\rm max}=7.5$
and
$P_{\rm min}\le P\le P_{\rm max}$.
The initial mass-ratio and period distribution are consistent with
pre-main sequence binary star data (K1). The initial eccentricity
distribution is assumed to be dynamically relaxed, $f_{\rm e}(e)=2\,e$
(equation~4 in K1), but is not critical here.

For each
binary and single-star cluster we perform $N_{\rm run}=5$ and~3
simulations, respectively. All results quoted here are averages of $N_{\rm
run}$ simulations.

In addition to the simulations of the above six clusters we perform $N_{\rm
run}=20$ simulations of the dominant mode cluster studied by K2 which has
$R_{0.5}=0.85$~pc initially. The
$N_{\rm
bin}=200$ binary systems per simulation have a mass-ratio distribution and a
birth eccentricity distribution as above, but a birth period distribution
given by equation~8 in K2. The birth orbital parameters of the short period
(log$_{10}P<2-3$) binary systems are assumed to eigenevolve during the
proto-stellar accretion phase on a timescale of approximately $10^5$~yrs (see
section~2 in K2). The resulting distribution of orbital elements is the initial
($t=0$)
distribution for the N-body simulation, and is consistent with the mass-ratio
and eccentricity distributions observed for short-period G~dwarfs, i.e. a bias
towards equal mass components and a bell shaped eccentricity distribution,
respectively. The minimum orbital period is about 3~days, and in our model
about 3~per cent of all systems are merged binaries at the start of the N-body
simulation.  The initial
period distribution can be approximated by $f_{\rm P,in}({\rm
log}_{10}P)=3.50\,{\rm log}_{10}P\,/[100+({\rm log}_{10}P)^2]$
(equation~11 in K1, see also fig.~7 in K2).
Stimulated evolution (i.e. the changes in orbital parameters due to
interactions with other systems) does not significantly change the
orbital parameters of binaries with log$_{10}P<3$.

Physical parameters for each cluster discussed here are listed in table~1
of K1, and in Table~1 we repeat some of these.

\bigbreak
\vskip 3mm
\bigbreak

\hang{ {\bf Table 1.} Initial stellar clusters}

\nobreak
\vskip 1mm
\nobreak
{\hsize 17 cm \settabs 21 \columns

\+& $R_{\rm 0.5}$ &$N_{\rm bin}$ &$N_{\rm sing}$ &~~$f_{\rm tot}$
&&$n$~~~log$_{10}(n_{\rm c})$& &&$\sigma$
&$T_{\rm cr}$ &&$T_{\rm relax}$\cr

\+ &pc & & &$t=0$
&&stars &stars  &&km
&Myrs &&Myrs\cr
\+&&&&&&pc$^{-3}$ &pc$^{-3}$&&sec$^{-1}$ \cr

\+ &0.08 &200 &~~0 &~~~1&&12 &5.6 &&1.7 &0.094 &&0.30\cr
\+ &0.25 &200 &~~0 &~~~1&&12 &4.1 &&0.9 &0.54  &&1.8 \cr
\+ &0.77 &200 &~~0 &~~~1&&11 &2.7 &&0.5 &3.0   &&9.5 \cr
\+ &2.53 &200 &~~0 &~~~1&&5  &1.1 &&0.3 &17    &&56  \cr
\+ &0.08 &0 &~~400 &~~~0&&12 &5.6 &&1.7 &0.094 &&0.30\cr
\+ &0.25 &0 &~~400 &~~~0&&12 &4.1 &&0.9 &0.54  &&1.8 \cr

\+ &0.85 &200 &~~0 &~~~1&&10 &2.5 &&0.5 &3.5   &&11  \cr
}
\vskip 2mm

\hang{$n_{\rm c}$ is the central number density; $\sigma$ the average velocity
dispersion; $T_{\rm cr}$ and $T_{\rm relax}$ are, respectively, the crossing
and relaxation times (for further details see section~3 in K1).}

\bigbreak\vskip 3mm

We refer to a single star as a single star system and define the
overall proportion
of binary systems at time $t$ to be $f_{\rm tot}(t)=N_{\rm bin}(t)\,/(N_{\rm
bin}(t)+N_{\rm sing}(t))$ (equation~2 in K1). This definition extends to
stellar systems in any sub-domain, e.g. if orbits only in a specific period
range are accessible (cf. with equation~7 in K1) or if $f_{\rm tot}(t)$ is
evaluated in a particular volume of space or mass range.

The simulations of the dynamical evolution of the above seven clusters allow an
intercomparison of the evolution of
overall properties of the stellar systems such as stellar number density, mean
stellar mass
and binary star segregation, and the distribution of centre-of-mass kinetic
energies (in the local rest frame).
To quantify the evolution of the clusters we measure the number density,
$n(t)$, mean stellar mass, ${\overline m}(t)$, and the overall proportion of
binaries,
$f_{\rm tot}(t)$, within a standard spherical volume, $V$, with radius $R$
centered on the number density maximum of the cluster. For example, if we take
$R=2$~pc then we refer to $V$ as the {\it central 2~pc sphere}. $f_{\rm
tot}(t)$
is defined above, and in this context $N_{\rm sing}(t)$ and $N_{\rm bin}(t)$
are the number of single and binary systems in $V$, respectively. Similarly,
${\overline
m}(t)=M_{\rm stars}(t)/(N_{\rm sing}(t)+2\,N_{\rm bin}(t))$, where $M_{\rm
stars}(t)$ is the total stellar
mass in $V$, and $n(t)=(N_{\rm sing}(t)+2\,N_{\rm bin}(t))/V$. The three
quantities
$n(t)$, ${\overline m}(t)$ and $f_{\rm tot}(t)$ are readily accessible to an
observer ($t=$cluster age). If an instrumental flux limit and/or a resolution
limit limits the observations then we can in principle apply the same bias to
our model
values $n(t)$, ${\overline m}(t)$ and proportion of binaries. This will be
necessary in a case-by-case treatment of individual Galactic clusters.

For conciseness we write $N_{\rm bin}=N_{\rm bin}(0)$, $N_{\rm sing}=N_{\rm
sing}(0)$, and $R_{0.5}=R_{0.5}(0)$ in the knowledge that the half mass radius
of a cluster evolves. We do not evaluate $R_{0.5}(t>0)$ because this is non
trivial requiring exact knowledge of escapers. In Section~3.1 we show that
all clusters with 400~stars disintegrate completely after 800~Myrs, and from
hereon we refer to the final
distributions of, say binary star binding energies, as the distributions
evaluated at $t=1$~Gyr that result after cluster disintegration.

\nobreak\vskip 10pt\nobreak
\noindent{\bf 3. CLUSTER EVOLUTION}
\nobreak\vskip 10pt\nobreak
\noindent
In this section we concentrate on the evolution of the number density,
$n(t)$, mean stellar
mass, ${\overline m}(t)$, and proportion of binaries, $f_{\rm tot}(t)$, in the
four stellar clusters with
$N_{\rm bin}=200$ and $R_{0.5}=0.077,0.25,0.77,2.53$~pc, and in the two
single-star clusters with $N_{\rm sing}=400$ and $R_{0.5}=0.077,0.25$~pc.

\nobreak\vskip 10pt\nobreak
\noindent{\bf 3.1 Lifetime and birthrate}
\nobreak\vskip 10pt\nobreak
\noindent
{}From Fig.~1 we infer
that there is no significant difference in the evolution of $n(t)$ between the
various clusters. After the first two relaxation times, $n(t)$ for the
$R_{0.5}=2.53$~pc cluster joins the other evolutionary curves and
the clusters depopulate at the same rate, irrespective of the
presence of primordial binaries (some of which are hard - see
Fig.~5 below) and of the initial cluster concentration. A reasonable
approximation of the depopulation of the central 2~pc sphere is

$$ {\rm log}_{10}n(t)=1.0-1.9\times10^{-3}\, t,   \eqno (1a)$$

\noindent where $t$ is in Myrs and $t<600$~Myrs. Equation~1a can be rewritten
to

$${n(t)\over10} = {\rm e}^{-{t\over \tau_{\rm e}}},  \eqno (1b)$$

\noindent where $\tau_{\rm e}\approx230$~Myrs is the exponential decay time.

We measure the disintegration time of each cluster
to be the time taken until
the number density has reached $0.1\,$stars$\,$pc$^{-3}$ (i.e. only
three stars remain in the central 2~pc sphere), which
is characteristic of the Galactic disk in the proximity of the Sun. The
mean life time, $\tau_{0.1}$, is thus a strict upper limit and is listed in
table~1 of K1.
{}From an observational point of view we might expect that a stellar cluster
with
about 100~stars in the central 2~pc sphere (i.e. $n(t)\approx3$
stars~pc$^{-3}$) might be closer to the detection limit of an open cluster. We
therefore also consider the alternative definition of the lifetime of a cluster
$n(\tau_{3.1})=3.1$ stars~pc$^{-3}$ (i.e. $1\,M_\odot$~pc$^{-3}$, c.f. Lada \&
Lada 1991). From equation~1 and Fig.~1 we obtain $\tau_{3.1}\approx250$~Myrs.

We conclude that clusters with $R_{0.5}\ge0.077\,$pc and initially with~400
stars have lifetimes $\tau_{0.1}\approx700$~Myrs and
$\tau_{3.1}\approx250$~Myrs. McMillan \& Hut (1994) also note that primordial
binaries do not significantly affect the cluster evaporation timescale.
The finding that $\tau_{0.1}$ and $\tau_{3.1}$ are
independent of $N_{\rm bin}, N_{\rm sing}$ (as long as $N_{\rm sing}+2\,N_{\rm
bin}=400$) and $R_{0.5}$
disproves the theory of cluster lifetimes based on the assumption of a
constant rate of evaporation from a cluster. Wielen (1988) shows that this
theory implies a constant lifetime for $R_{0.5}>0.2$~pc, and a steep
decay of the lifetime for $R_{0.5}<0.2$~pc.

Our $\tau_{0.1}$ is about 70~times
as large as the lifetime of real clusters found by Battinelli \&
Capuzzo-Dolcetta (1991). In part this is due to our $\tau_{0.1}$ being a strict
upper
limit whereas the lifetime discussed by Battinelli \& Capuzzo-Dolcetta (1991)
refers to the time it takes for half of all clusters to be destroyed, but our
result does confirm that other mechanisms than internal dynamical evolution
must be responsible for cluster disintegration. For example, encounters with
giant molecular
clouds are destructive (Terlevich 1987) and
can reduce cluster lifetimes on average to approximately $100$~Myrs (Theuns
1992).

Little is known about the typical star-forming systems that eventually lead to
the stellar population in the Galactic disk. Our inverse dynamical population
synthesis in K1 indicates that the majority of Galactic field stars may
result from clustered star formation. However,
Wielen (1971) finds that only a few per cent of the Galactic disk stars are
born in Galactic clusters by showing that there are too few Galactic clusters
in total and that the lifetimes of the clusters are too long. Thus, if
clustered star formation is the dominant mode then most birth
clusters of young stellar objects are not bound gravitationally after dispersal
of the remnant cloud material. Conversely, the Galactic clusters must remain
gravitationally bound and probably result from
rather rare incidences of high local star formation efficiency. These issues
are discussed in greater detail by Lada, Margulis \& Dearborn (1984), Mathieu
(1986), Pinto (1987) and Verschueren \& David (1989).

If the majority of low mass stars in the Galactic disc
are born in initially unbound clusters of, say 200~binary systems, then we
require a cluster birth rate of approximately
$15\,$clusters$\,$kpc$^{-2}$Myr$^{-1}$, assuming a constant birth
rate over $10^{10}\,$yr, a vertical Galactic disc scale height of $300\,$pc
and a local stellar number density of $0.1\,$pc$^{-3}$. This is
approximately 30~times the birth
rate for the Galactic clusters as obtained by Battinelli et al.
(1994). If the majority of stars form in embedded clusters then their
sample cannot be complete. The observational catalogues probably do not
account for the total birth sample which
includes initially embedded and later unbound aggregates of a few
hundred low mass pre-main sequence stars.

\nobreak\vskip 10pt\nobreak
\noindent{\bf 3.2 Mass and binary star segregation}
\nobreak\vskip 10pt\nobreak
\noindent
For a given Galactic cluster we are
unable to observe the entire birth population, but we
can obtain data from
the central region quite readily. The dynamical properties
of stellar systems in the
central region of a cluster is determined by its past dynamical evolution.

{}From Fig.~2 it is evident that the mean stellar mass
within the central 2~pc sphere increases linearly until
disintegration time when  ${\overline m}(t)\approx
0.45\,M_\odot$, whereas outside it is
$0.32\,M_\odot$. Mass segregation does not strongly depend on the initial
cluster size (see also Fig.~4 below). Writing

$${\overline m}(t) = 0.32\,M_\odot + \mu\,t,    \eqno (2)$$

\noindent with $t<600$~Myrs, we obtain
$\mu=0.28\times10^{-3}\,M_\odot$~Myr$^{-1}$ ($R_{0.5}=2.53$~pc),
$\mu=0.23\times10^{-3}\,M_\odot$~Myr$^{-1}$ ($R_{0.5}=0.77$~pc),
$\mu=0.22\times10^{-3}\,M_\odot$~Myr$^{-1}$ ($R_{0.5}=0.25$~pc),
$\mu=0.20\times10^{-3}\,M_\odot$~Myr$^{-1}$ ($R_{0.5}=0.08$~pc).
However, mass segregation is significantly more
pronounced in the single-star clusters where we obtain
$\mu\approx0.37\times10^{-3}\,M_\odot$~Myr$^{-1}$.
This results because the mean mass per centre-of-mass particle
is larger in the binary star clusters than in the single star clusters. In all
panels of Fig.~2 the mean stellar mass outside the central 2~pc sphere
decreases within the first 50~Myrs because low-mass stars are ejected
preferentially. The `escaped' low mass stars
have a smaller mean stellar mass in the single-star clusters because in
the binary star clusters the least massive stars
are initially bound in binary systems. Some of these will form a halo
population being unable to find an exit in the equipotential surface of the
remnant cluster plus Galaxy (see e.g. Terlevich 1987).

In fig.~3 of K1 we show that the overall proportion of binary systems
$f_{\rm tot}(t)$ (counting
all cluster members and non-members) is depleted at a rate which is a sensitive
function of the initial concentration.
In Fig.~3 we see that the overall
proportion of binaries within the central 2~pc sphere, $f_{\rm in}(t)$, is
significantly larger
than the proportion of binaries outside, $f_{\rm out}(t)$. The spatial
segregation
of binary systems results because their mean mass is larger than that of single
stars. After the initially rapid ionisation of the wide binaries, $f_{\rm in}$
increases only by a small amount as the cluster ages.
This result is also obtained by McMillan \& Hut (1994). For the four
clusters we define the ratio $r_1(R_{0.5}) = f_{\rm out}/f_{\rm in}$  and
$r_2(R_{0.5})=f_{\rm tot}^{\rm obs}/f_{\rm in}$ at approximately 600~Myrs. In
the Galactic field $f_{\rm tot}^{\rm obs}=0.47\pm0.05$ (K1, K2). The ratios
$r_1$ and $r_2$ are tabulated in Table~2.

\bigbreak
\vskip 3mm
\bigbreak

\hang{ {\bf Table 2.} Binary stars in the central 2~pc sphere at $t=600$~Myrs}

\nobreak
\vskip 1mm
\nobreak
{\hsize 17 cm \settabs 21 \columns

\+$R_{0.5}$ &&&$r_1$ &&&$r_2$\cr
\+pc        \cr

\+0.08 &&&0.53 &&&0.93\cr
\+0.25 &&&0.63 &&&0.74\cr
\+0.77 &&&0.75 &&&0.58\cr
\+2.53 &&&1.0  &&&0.57\cr
}
\bigbreak\vskip 3mm

Table~2 illustrates that the proportion of binaries outside the 2~pc sphere
drops significantly with decreasing initial cluster size ($r_1$). The
proportion of binaries in the central 2~pc sphere can, however, be similar in
an initially highly concentrated cluster to that observed in the field ($r_2$)
(compare with the Trapezium Cluster in Section~6.3). This population of binary
stars is depleted significantly at log$_{10}P\ge5$
when compared to the Galactic field population (see fig.~5 in K1).

\nobreak\vskip 10pt\nobreak
\noindent{\bf 4 THE DOMINANT MODE CLUSTER AND DYNAMICAL EQUIVALENCE}
\nobreak\vskip 10pt\nobreak
\noindent
In this section we study the evolution of number density, mass segregation and
binary star proportion in the dominant mode cluster $\left[(N_{\rm
bin},R_{0.5})=(200,0.85\,{\rm pc})\right]$, and we investigate which
combination of $R_{0.5}$ and $N_{\rm bin}$ might define clusters that are
dynamically equivalent to the dominant mode cluster.

\nobreak\vskip 10pt\nobreak
\noindent{\bf 4.1 Evolution of the Dominant Mode Cluster}
\nobreak\vskip 10pt\nobreak
\noindent
In the top panel of Fig.~4 we show the average number density
evolution, $n(t)$. It does not differ from the evolution of the six
clusters discussed above, confirming our conclusion that after an
initial relaxation phase, $n(t)$ does not depend on initial values of
$R_{0.5}$, log$_{10}P_{\rm min}$ or $f_{\rm tot}$. Our dominant mode
cluster has a lifetime $\tau_{0.1}=740\pm150$~Myrs, as obtained from
the 20~individual simulations.

Mass segregation proceeds similarly to the six clusters discussed in Section~3
(Fig.~4). The initial mean stellar mass,
${\overline m}(0)$, is somewhat larger in the present simulations because a few
per cent of all binaries have merged to single stars during pre-main
sequence eigenevolution (K2). However, the slope
of the approximately linear ${\overline m}(t)$ relation is about the same as
for our
$R_{0.5}=0.77$~pc cluster (equation~2), and is smaller than for the single-star
clusters.

The evolution of the overall proportion of binaries, $f_{\rm tot}(t)$, inside
and outside the 2~pc sphere is shown in the bottom panel of Fig.~4. As found in
Section~3, $f_{\rm in}(t)$ first decreases rapidly to a minimum ($f_{\rm
in}\approx0.6$) after about 10 initial relaxation times and then rises slowly.
Also, $r_1=0.73\approx r_2=0.71$ as expected for the dominant mode cluster
(compare with Table~2). However, $f_{\rm
in}(R_{0.5}=0.85\,{\rm pc})<f_{\rm in}(R_{0.5}=0.77\,{\rm pc})$ because our
present initial binary star population extends to larger periods than in the
$R_{0.5}=0.77$~pc simulation discussed in Section~3. The large proportion of
binaries in the central cluster region is interesting in the context of
the finding by Kroupa \&
Tout (1992) that a large proportion of binary systems is consistent with
the distribution of data in the colour--magnitude diagram of the Praesepe
Cluster which has an age of about $8\times10^8$~yrs (Cayrel de Strobel 1990).

Fig.~4 demonstrates the following points: (i) $n(t)$ is invariant to
the initial conditions; (ii) ${\overline m}(t)$ increases approximately
linearly with time at a rate $\mu$ that depends primarily on whether the
cluster is composed
initially of binaries or single stars and only secondarily on $R_{0.5}$, and a
zero point, ${\overline
m}(0)$, that depends on the stellar mass function and on whether some of the
birth binaries merge to more massive single stars; and (iii)
$f_{\rm in}(t)$
is approximately constant, but depends sensitively on the initial $R_{0.5}$ and
on the initial distribution of periods.

\nobreak\vskip 10pt\nobreak
\noindent{\bf 4.2 Dynamically equivalent clusters?}
\nobreak\vskip 10pt\nobreak
\noindent
We define a stellar aggregate or cluster, which is {\it dynamically equivalent}
to the dominant mode cluster, to be an aggregate or cluster initially not
necessarily in virial equilibrium with $R_{0.5}'\neq R_{0.5}=0.85$~pc and
$N_{\rm bin}'\neq N_{\rm bin}=200$, in which the initial dynamical properties
of stellar systems, as defined in Section~2, evolve to distributions after
cluster disintegration, that are consistent with the observed distributions in
the Galactic field.

Assuming the dynamical properties of stellar systems at birth are invariant to
the initial conditions, we
postulate that $n(t)$, ${\overline m}(t)$ and $f_{\rm in}(t)$ uniquely
specify the evolutionary path of a stellar cluster in virial equilibrium. That
is, if we can measure
these quantities for some Galactic cluster then we can probably uniquely
specify the initial conditions $(N_{\rm bin},R_{0.5})$. In Section~3 we have
shown that after an
initial relaxation phase, $n(t)$ is independent of the initial $R_{0.5}$. From
Wielen (1988, fig.~2) and Terlevich (1987, fig.~3) we find that
$\tau_l$ and $n(t)$ can be scaled to any initial
$n'(0)$ (i.e. $N_{\rm bin}'$) by a multiplicative factor:

$$ \tau_l' = {N_{\rm bin}' \over N_{\rm bin}}\, \tau_l  \eqno
(3a)$$

\noindent and

$$n'(t) = {N_{\rm bin}' \over N_{\rm bin}}\, n(t),  \eqno (3b)$$

\noindent where $\tau_l$ is $\tau_{0.1}$ or $\tau_{3.1}$. The
exponential decay time $\tau_{\rm e}$ is invariant to changes in initial
conditions.
The evolution of ${\overline m}(t)$ and $f_{\rm tot}(t)$,
however, depends on $R_{0.5}$.

We probably obtain the same overall stimulated
evolution of orbital parameters if $n_{\rm c}(0)\,\sigma(0)\approx160$~stars
pc$^{-2}$ Myr$^{-1}$, where $n_{\rm c}(0)$ and $\sigma(0)$ are the initial
central number density and average velocity dispersion, respectively (Table~1).
This implies $R_{0.5}\propto N_{\rm bin}^{3 \over 7}$, i.e.

$$ R_{0.5}' \approx 0.088\,N_{\rm bin}'^{3\over7}\,{\rm pc}, \eqno (4)$$

\noindent for aggregates that are dynamically equivalent to our dominant mode
cluster. Alternatively, if we take the ratio of $\tau_{0.1}/T_{\rm relax}$ to
be an estimate of the number of relaxations the system undergoes in its
lifetime $\tau_{0.1}$ then our dominant mode cluster is characterised by
$\tau_{0.1}/T_{\rm relax}\approx70$. For constant $\tau_{0.1}/T_{\rm relax}$ we
obtain $R_{0.5}\propto N_{\rm bin}^{1\over3}\left[{\rm log}_{10}(0.8\,N_{\rm
bin})\right]^{2\over3}$, i.e.

$$R_{0.5}'\approx 0.086\,N_{\rm bin}'^{1\over3}\left[{\rm log}_{10}(0.8\,N_{\rm
bin}')\right]^{2\over3}\,{\rm pc}, \eqno (5)$$

\noindent using the definition for $T_{\rm relax}$ (K1) and the scaling
property of $\tau_{0.1}$ above.

Simulations of clusters with $R_{0.5}'$ and $N_{\rm bin}'$ scaled according to
equations~4 and~5 are necessary to verify our postulate and assertions.
Additional simulations
including stars more massive than $1.1\,M_\odot$ and
other initial configurations, such
as subclustering, cold collapse and/or changing background potential owing to
gas removal, which must be important in the first 10~Myrs of cluster evolution,
will be necessary for a comparison with real open clusters.
Furthermore, the evolution of real Galactic clusters can be affected by
passing interstellar clouds (Terlevich 1987, Theuns 1992), which ought to be
kept in mind when comparing Galactic clusters with models.

While we can apply the concept of dynamical equivalence to embedded clusters
that disperse within about $10^7$~yrs by our definition in the first paragraph
of this section, we cannot apply equations 1--5 to these (see also section~6.4
in K1).

\nobreak\vskip 10pt\nobreak
\noindent{\bf 5 THE KINEMATICAL SIGNATURE OF STAR FORMATION}
\nobreak\vskip 10pt\nobreak
\noindent
Past work (Heggie 1975, Hills 1975) has etablished that binary systems
are ionised at a rate which is a function of the ratio of the internal binding
energy, $-E_{\rm bin}$, of the binary and the centre of mass kinetic energy,
$E_{\rm kin}$, or temperature of the surrounding field population. The
hardening of relatively hard ($E_{\rm bin}/E_{\rm kin}>1$) binaries heats
the field thereby increasing the number of systems with relatively large
$E_{\rm kin}$. In this section we study the observational consequences of these
processes. Our aim is to understand the evolution of the binary star binding
energy distribution, $f_{E_{\rm bin}}$, and of the centre of mass kinetic
energy distribution, $f_{E_{\rm kin}}$, as the clusters evolve. We then apply
the insights gained to star forming aggregates by invoking dynamical
equivalence.

\nobreak\vskip 10pt\nobreak
\noindent{\bf 5.1 Energetics}
\nobreak\vskip 10pt\nobreak
\noindent
For each of the four binary star clusters
discussed in Section~3
we compute $f_{E_{\rm bin}}$ and $f_{E_{\rm kin}}$.
For comparison we also compute $f_{E_{\rm kin}}$
for the single-star clusters discussed in Section~3. The final
$f_{E_{\rm kin}}$ are tabulated in Table~A-1.
In Fig.~5 we show the initial
distributions and the final distributions after cluster dissolution.
Units of energy are
$M_\odot\,$km$^2$/sec$^2$ (i.e. in units of
$1.99\times10^{43}$erg).

The initial $f_{E_{\rm bin}}$
are equal in all four binary star clusters, but the
initial kinetic energies increase with decreasing cluster radius. The
increasing overlap of the initial $f_{E_{\rm bin}}$ and $f_{E_{\rm kin}}$
with decreasing cluster size, and the
increased number density, leads to
an increasingly efficient destruction rate of binary systems.
A substantial number of initial binaries are hard (i.e. have $E_{\rm
bin}/E_{\rm kin}^{\rm max}>1$), even in our most compact initial cluster
($R_{0.5}=0.08$~pc, $E_{\rm kin}^{\rm max}\approx4\,M_\odot$~km$^2$ sec$^{-2}$)
and even though
$P_{\rm min}>2.7$~yrs. The final $f_{E_{\rm kin}}$ in
the binary star clusters is significantly enhanced at log$_{10}E_{\rm kin}>-1$
when
compared to the single star clusters, which have initially the same $R_{0.5}$
($0.25$ and 0.08~pc).
Our most compact binary star cluster with
$R_{0.5}=0.08\,$pc has a final high kinetic energy tail that is virtually
identical to the high binding energy tail (Fig.~6). We also observe in Fig.~5
that increasingly
harder binaries appear with decreasing $R_{0.5}$, being
the result of enhanced stimulated evolution per unit time in a cluster with
smaller initial crossing time.

The final high kinetic and binding energy
tails come about because binaries with $E_{\rm bin}>E_{\rm kin}^{\rm
max}$ on average acquire additional
binding energy after interaction with a field particle
and are not ionised (Heggie 1975). The perturbing star or system involved
in the energy
exchange gains kinetic energy. Energy exchange proceeds until the cluster is
depopulated. While the average lifetime of the open cluster is not
affected (Section~3.1), the final $f_{E_{\rm bin}}$
and $f_{E_{\rm kin}}$ contain information on the initial distribution of
orbital periods, the initial velocity dispersion and the initial number
density.
Using the data in Table~A-1 we compare in
Table~3 for
each cluster the final proportion of systems with log$_{10}E_{\rm kin}\ge1$.

\bigbreak
\vskip 3mm
\bigbreak

\hang{ {\bf Table 3.} Fraction, $g_{E_{\rm kin}}$ in per cent, of
center-of-mass systems with $E_{\rm kin}\ge10\,M_\odot$~km$^2$~sec$^{-2}$

\nobreak
\vskip 1mm
\nobreak
{\hsize 17 cm \settabs 21 \columns

\+$R_{0.5}$ &&&&&$g_{E_{\rm kin}}$\cr
\+~pc       &&$[f_{\rm tot}(t=0)=1]$ &&&$[f_{\rm tot}(t=0)=0]$ \cr

\+0.08 &&&8.0 &&&2.2\cr
\+0.25 &&&4.6 &&&1.1\cr
\+0.77 &&&2.0 &&&--\cr
\+2.53 &&&0.2 &&&--\cr
}
\bigbreak\vskip 3mm

In K2 we study in detail the distribution of dynamical properties of stellar
systems
that result if the majority of stars in the Galactic disk are born in the
dominant mode cluster specified by $(N_{\rm bin},R_{0.5})=(200,0.85\,{\rm
pc})$. The distribution of binary star periods we adopt in
K2 is more realistic than the flat distribution assumed for
the four binary star clusters above, because it spans the entire observed range
from log$_{10}P_{\rm min}\approx0.5$ to log$_{10}P\approx9$.
It is useful to compare this
larger set of simulations of the dominant mode cluster with the results of the
$(N_{\rm bin},R_{0.5})=(200,0.77\,{\rm pc})$ cluster discussed above, and to
extend our analysis to the observational plane, i.e. to study the distribution
of velocities of centre-of-mass systems.

In Fig.~7 we compare the initial and final $f_{E_{\rm kin}}$
and $f_{E_{\rm bin}}$. The former are virtually
identical while the latter differ significantly. The initial $f_{E_{\rm bin}}$
extends to much higher values in the realistic case. These
hard binaries, initially with $E_{\rm bin}/E_{\rm kin}^{\rm max}>25$ ($E_{\rm
kin}^{\rm max}\approx1\,M_\odot$~km$^2$ sec$^{-2}$), do not
affect the dynamics of the cluster because their interaction cross sections are
too small to be significant. The initially larger number of binaries in the
$R_{0.5}=0.77$~pc cluster, with $0.04\le E_{\rm bin}/E_{\rm kin}^{\rm max} \le
25$, do not significantly change the final distribution of $E_{\rm kin}$, when
compared with the $R_{0.5}=0.85$~pc cluster (top panel of Fig.~7), because the
clusters do not survive long enough to exhaust this binary star
population. At cluster dissolution time a sufficient number of these binaries
survives in both clusters. Nevertheless, it is these binaries which heat the
cluster leading to the final high $E_{\rm kin}$ tail apparent in the top panel
of Fig.~7. Finally, the binaries with $E_{\rm bin}/E_{\rm kin}^{\rm max}<0.04$
are ineffective energy sinks and are ionised without affecting the temperature.
Thus, the different form of the
initial $f_{E_{\rm bin}}$ in the $R_{0.5}=0.85$~pc cluster
does not change the kinematical signature of star formation, which we now
consider in greater detail.

In Fig.~8 we illustrate the changes in the $E_{\rm bin}$ and $E_{\rm kin}$
distributions that result after cluster dissolution. We observe a significant
increase in the number of systems with
log$_{10}E_{\rm kin}\le -1.4$, which comes about because these systems had to
overcome
the cluster potential. Including gas removal during early cluster evolution
would reduce this gain. There is also a significant increase in the
number of systems with $E_{\rm kin}>0.32\,M_\odot$ km$^2$ sec$^{-2}$
which is due to binary star heating. The
maximum final kinetic energy that results in our model is approximately
$3.2\times10^3\,M_\odot$ km$^2$ sec$^{-2}$.

\nobreak\vskip 10pt\nobreak
\noindent{\bf 5.2 Kinematics}
\nobreak\vskip 10pt\nobreak
\noindent
The interaction of multiple star systems, ionisation of binaries and the
disintegration of a cluster changes the distribution of velocities of the
stars.
High velocity escapers, with velocities larger than $10\,$km~sec$^{-1}$ and up
to 100~km~sec$^{-1}$ or larger (see Leonard \& Duncan 1990), are the result of
the internal dynamics of
stellar systems. The distribution of velocities will be different for clusters
consisting initially only of single stars than if these are composed of a large
proportion of primordial binaries (Fig.~5). In this section we consider the
$N_{\rm run}=20$ simulations of the dominant mode cluster $(N_{\rm
bin},R_{0.5})=(200,0.85\,{\rm pc})$.

In the top panel of Fig.~9 we show the initial distribution of centre of mass
velocities. The dashed histogram shows the distribution in a
Plummer sphere in virial equilibrium with velocity dipersion
$\sigma\approx0.5\,$km~sec$^{-1}$. The mean system mass at time $t=0$, shown as
open circles in the middle panel of Fig.~9, is $0.65\,M_\odot$. It is
independent of velocity and has the value expected for random
pairing from our mass function (equation~1 in K1).
After cluster disintegration we are
left with essentially the same low-velocity tail of the centre of mass velocity
distribution (solid histogram in the top panel of Fig.~9; note the
distributions are normalised to unit area so that the increased
number of systems with small velocity owing to cluster dissolution (see
Figs.~7 and~8) is thus not apparent here). However, there is
now an appreciable high velocity tail. About 1.5~per~cent of all systems have a
velocity greater than 10~km~sec$^{-1}$. The mean system mass decreases
systematically with increasing velocity up to about 1--2~km~sec$^{-1}$
(filled circles in the middle panel of Fig.~9) reflecting the
establishment of equipartition of energy in a quasi-equilibrium system.
For larger velocities there is no correlation because of the
stochastic process of near encounters. The mean mass of stars
in the second highest velocity bin is $0.82\,M_\odot$, whereas the fastest star
ejected from the cluster has a velocity of about 70~km~sec$^{-1}$ and a mass of
$0.18\,M_\odot$. Thus the change in character of the final velocity data at
about
1--2~km~sec$^{-1}$, which approximately corresponds to the maximum kinetic
energy available in the initial cluster (Figs.~7 and~8 below), is a result of
the process of equipartition of kinetic energy being replaced by the
stochastic shedding
of energy from relatively hard binaries to the field population.

The bottom panel of Fig.~9 shows the number of single stars and binaries
initially and after cluster dissolution as a function of centre of mass
velocity. After cluster disintegration
most of the high velocity systems are single stars, with the two highest
velocity bins containing no binaries. The small number of single stars at $t=0$
results from merging during pre-main sequence eigenevolution (see K2) and wide
pairs being ionised immediately in the crowded central region
of the cluster, where we have about 320~stars~pc$^{-3}$ (Table~1).

\nobreak\vskip 10pt\nobreak
\noindent{\bf 5.3 Discussion}
\nobreak\vskip 10pt\nobreak
\noindent
Kinematical data of all stars of equal age in the vicinity of a star
forming region should reveal
similar distributions if the present scenario is correct, although we must keep
in mind the bias introduced here by neglecting massive stars,
stellar evolution effects and a changing background potential.

Including massive stars
will probably raise the maximum velocity of ejected stars because more binding
energy is available in massive binaries. Leonard \& Duncan (1990) obtain
runaway stars with velocities up to about 200~km~sec$^{-1}$ for
$5\,M_\odot$ stars and up to about $50\,$km sec$^{-1}$ for
$20\,M_\odot$ stars corresponding to log$_{10}E_{\rm kin}=5$ and
log$_{10}E_{\rm kin}=4.4$, respectively.
Stellar evolution effects tend to unbind the cluster reducing the
work to be done when leaving the cluster potential, and would appear to be
effective in Galactic clusters rather than embedded clusters, which have ages
less than about 1--10~Myrs.
Also, the initial dynamical evolution of a cluster of proto-stars may be
dominated by a cold collapse rather than our assumed virial equilibrium.
Aarseth \& Hills (1972) demonstrate that higher ejection velocities are
achieved in this case, but that the cluster relaxes within about two
free-fall times to a configuration it would have had if it had been in virial
equilibrium initially. In this case we would expect a larger number of high
velocity escapers,
but not a significantly different distribution of final kinetic energies.
The overall distribution of $E_{\rm
kin}$ should not change significantly if stars more massive than $1.1\,M_\odot$
are included, because of the steeply decreasing initial mass function with
increasing stellar mass.

A much more significant
effect is the expulsion of a significant amount of binding mass during the
first 10~Myrs (Mathieu 1986, see also section~6 in K1). This
would imply that the young cluster expands without having to overcome its own
binding energy thereby more or less freezing $f_{E_{\rm kin}}$, as discussed in
greater detail by Verschueren \& David (1989, see also Pinto 1987).

Given our results in K1, we now postulate that most stars form in aggregates
that are dynamically equivalent
to the dominant mode cluster, and we contemplate the implications this has for
the kinematics and distribution of young stars. The mass in stars is $M_{\rm
stars}=\epsilon\,M_{\rm tot}(0)$, where $M_{\rm tot}(0)$ is the total initial
mass of the star-forming core. Observations indicate that the star formation
efficiency $\epsilon\approx0.1-0.3$ (see e.g. Mathieu 1986) so that virial
velocities of stellar systems in a very young aggregate will be dictated by the
mass remaining in the gas, $M_{\rm gas}(t)=M_{\rm tot}(t)-M_{\rm stars}$. The
initial velocity dispersion for such a system is $\sigma=0.042(M_{\rm
tot}(0)/R_{0.5})^{1\over2}$~km sec$^{-1}$ (assuming virial equilibrium and a
Plummer density profile), where $M_{\rm tot}(0)$ is in $M_\odot$ and $R_{0.5}$
is in pc. Thus, if $\epsilon=0.1$, $R_{0.5}=0.5$~pc and $M_{\rm
stars}=128\,M_\odot$, then $\sigma=2.1$~km sec$^{-1}$. Assuming a mean system
mass of $0.64\,M_\odot$ then about 95~per cent of all systems will have $E_{\rm
kin}<6\,M_\odot$~km$^2$ sec$^{-2}$.
After 10~Myrs most stars will
retain $E_{\rm kin}<6\,M_\odot$ km$^2$ sec$^{-2}$ irrespective of how rapidly
$M_{\rm gas}(t)$ tends to zero.
Dynamical equivalence implies that the number of systems with $E_{\rm
kin}>10\,M_\odot$~km$^2$ sec$^{-1}$ will probably not be significantly
different to the value given in Table~3. Bearing in mind possible cold collapse
and massive stars, we roughly estimate $g_{E_{\rm kin}}(E_{\rm
kin}>10\,M_\odot\,{\rm km}^2\,{\rm sec}^{-2}) \approx 5$ per cent.

A kinetic energy log$_{10}E_{\rm kin}=1$
corresponds to the potential energy a young stellar
system of mass $0.32\,M_\odot$ has at the edge of a spherical giant molecular
cloud that has a typical mass of $1.5\times10^5\,M_\odot$ and diameter of 40~pc
(Blitz 1993). At a distance of 8~kpc from the Galactic
center, and assuming a mass of $10^{11}\,M_\odot$ within this distance,
the tidal radius of the molecular cloud would be roughly 60~pc.
This crude
estimate suggests that the great majority of young stars may remain in the
vicinity
of a molecular cloud for its entire life time (a few $10^7$~yrs) even though
the high kinetic energy tail is enhanced by binary star heating. The molecular
cloud will probably have a halo of young stars which have been ejected with
velocities near to the escape velocity at formation site.
They might be on orbits which are either bound to the molecular
cloud, or they might have escape energy with escape delayed significantly until
the stars ``find an exit'' in the openings of the equipotential surface
formed by the cloud and the Galaxy (see e.g. Terlevich 1987).
The presumed halo population will have had close encounters with relatively
hard binary systems and
will probably have lost much of the circumstellar material. The halo of young
stars is thus expected to be enriched with wTTS (weak-lined T-Tauri stars).
Given our simulations we expect of order of 5~per cent of all stars formed in
the molecular cloud to be deposited in this halo, but the details depend on the
mass, shape and extend of the molecular cloud, and on the dynamics of the
dominant mode embedded cluster (see also section~6 in K1).
Of particular importance in this context is the detection of wTTS
over a wide area surrounding the Orion star-forming region by Sterzik et
al. (1995).

A kinetic energy of $E_{\rm kin}=10$~km$^2$ sec$^{-2}$ corresponds to a
velocity $v=8$~km sec$^{-1}$ for a star of mass $0.32\,M_\odot$. Systems
with $v<5$ km sec$^{-1}$ are less likely to escape the molecular cloud complex
and systems with $v>5$ km sec$^{-1}$ have gained kinetic energy because of
binary star heating. From our data (Table~A-2)
we find that about 95~per cent of all
systems have a velocity $v<5$ km sec$^{-1}$ with respect to the star formation
site. Bearing in mind a possible cold collapse and massive stars we roughly
estimate that 90~per cent of all young systems are trapped in the molecular
cloud complex.
These systems, of which 50~per cent
are binaries, may diffuse over a region 25~pc in radius over a time span of
5~Myrs. A young star may acquire additional kinetic energy if it falls towards
the potential minimum of the molecular cloud and may cover a larger distance in
5~Myrs, and the formation site may have an additional velocity of a few
km~sec$^{-1}$ with respect to the centre of mass of the molecular cloud.
The remaining
5--10~per cent of all systems with velocities larger
than 5~km~sec$^{-1}$ have a binary proportion of about 15~per cent and will
spread throughout the molecular cloud, with only roughly 3~per cent of all
systems having large enough velocities ($v>8$~km sec$^{-1}$) to leave the cloud
altogether.

This entire population may resemble the 5--7~Myrs old distributed population
found by
Strom et al. (1993) in the L1641 molecular cloud, and may not necessarily imply
distributed star formation (see also Section~1). Given our reasoning here, such
a distributed
population is consistent with star formation proceeding in that cloud for about
7~Myrs with stars currently being born in the much younger aggregates. The
first embedded dominant mode clusters may have dissolved by now.
The lack of an apparent distributed population in the L1630 molecular cloud
(Lada \& Lada 1991) may simply be due to star formation beginning recently
in that cloud with not enough time being available for a distributed
population to be established.

Whether these considerations are correct can be determined observationally. By
measuring the proportion of binaries in the distributed population of L1641,
and their
distribution of periods, we can determine whether these have passed through the
dynamics of an aggregate dynamically equivlant to the dominant mode
cluster. That is, the distributed
population ought to have a proportion of about 50~per cent binaries, and a
period distribution similar to the main sequence period distribution.
The distributed population observed in Taurus--Auriga, on the other hand,
must have been born in an isolated star formation mode, because the observed
distribution of periods for these young systems appears to be unevolved (see
sections~1 and~2 in K1). Precise radial velocity and proper motion measurements
of a large sample of young stars in the vicinity of a molecular cloud will be
very important to quantify the kinematics.

\nobreak\vskip 10pt\nobreak
\noindent{\bf 6 EXAMPLES: HYADES, PLEIADES AND TRAPEZIUM CLUSTERS}
\nobreak\vskip 10pt\nobreak
\noindent
As suggested in Section~4.2 (see also Section~2) and assuming
the dynamical properties of stellar systems are invariant to the star
forming conditions in the Galactic disk,
the set of observables $n(t)$, ${\overline m}(t)$ and $f_{\rm
tot}(t)$ probably uniquely define the past dynamical evolution of the
cluster.

We consider the Hyades, Pleiades and Trapezium
Clusters because these have been observed extensively.
Unfortunately we do not have a complete census of all cluster members, nor do
we have a complete census of all binary systems in these clusters.
A very detailed and insightful investigation of the dynamics of the Pleiades
Cluster is provided by Limber (1962a, 1962b). Given the lack of observational
data, Limber neglects binary systems and has to extrapolate to stars fainter
than $M_{\rm V}\approx10$. Limber (1962b) argues that the cluster has relaxed
sufficiently so that the massive stars have settled near the cluster center.
Their present spatial distribution thus needs not reflect the birth
configuration. The binary
proportion in the Pleiades Cluster is 13~per cent for spectroscopic binaries
with approximately $3.3<M_{\rm V}<6.3$
($P<10^3$~days, Mermilliod et al. 1992), and about 46~per cent for photometric
low-mass binaries (Steele \& Jameson 1995).
For the Hyades Cluster, Griffin et al. (1988)
estimate that 30~per cent of all cluster members with $2.6<M_{\rm V}<10.6$
are radial velocity binaries.
For systems brighter than $M_{\rm V}\approx13$, Eggen (1993) finds a
photometric
binary proportion $f_{\rm tot}^{\rm phot}\approx0.4$. These binaries are
concentrated towards the cluster center and their proportion is higher than in
the field (compare with Figs.~1 and~4).
$f_{\rm tot}$ is thus likely to be significantly
larger in both clusters (Kroupa \& Tout 1992).
The best
determined observational quantity, however, is the low-mass stellar luminosity
function
for the Hyades and Pleiades Clusters, which contains information on all three
observables above.

In this section we consider the $N_{\rm run}=20$ simulations of the dominant
mode cluster $(N_{\rm bin},R_{0.5})=(200,0.85\,{\rm pc})$. We convert stellar
masses to absolute magnitudes using
the mass--luminosity relation derived and tabulated in Kroupa et al. (1993),
and obtain I- and K-band absolute magnitudes as in Kroupa (1995c).
The single star luminosity function, $\Psi_{\rm sing}$, is obtained by binning
all individual stars
into magnitude bins, and the system luminosity function, $\Psi_{\rm sys}$, is
obtained by binning
all single star systems and all binary systems (which are assumed to be
unresolved) into magnitude bins.

\nobreak\vskip 10pt\nobreak
\noindent{\bf 6.1 The theoretical luminosity function}
\nobreak\vskip 10pt\nobreak
\noindent
In this section we discuss the various features of the stellar luminosity
function.

In the upper panel of Fig.~10 we plot the K-band luminosity functions for all
individual stars and systems. It is
evident from this figure, that after cluster dissolution, the surviving
unresolved binary stars, which have
a period distribution as shown in fig.~7 in K2 and a mass ratio distribution
plotted in fig.~12 in K2, lead to a decay of the field star
luminosity function at $M_{\rm K}>7$. This is exactly the effect one observes
in the Galactic field (Kroupa et al. 1993, Kroupa 1995c).
The flattening of the luminosity function at
$M_{\rm K}\approx4.4$ ($M_{\rm V}\approx7$) is the `H$^-$ plateau', and the
maximum at $M_{\rm K}\approx7$ ($M_{\rm V}\approx12$) is the `H$_2$--convection
peak' (Kroupa, Tout \& Gilmore 1990). The model Galactic field star luminosity
functions are tabulated in table~2 in Kroupa (1995c).

In the lower panel of Fig.~10 we compare the model field star luminosity
functions, shown in the upper panel, with the single star
and the system luminosity functions inside the
central 2~pc sphere of the dynamically highly evolved dominant mode cluster
at age $t=480\,$Myr (i.e. after~44 initial relaxation times).

The luminosity function of all individual stars in the
central cluster region
is highly depleted in low mass stars owing to advanced mass segregation and
evaporation of stars.
The system luminosity function in the central cluster region is highly biased
towards bright systems
when compared to the field system luminosity function.

Consider the ratio which
is independent of the initial $N_{\rm bin}$ (i.e. the initial
number of stars in the cluster):

$$\zeta = { \Psi_{{\rm H}_2} \over \Psi_{{\rm H}^-} },
\eqno(6)$$

\noindent where $\Psi_{{\rm H}_2}$ and $\Psi_{{\rm H}^-}$ are the luminosity
functions at the H$_2$--convection peak and the H$^-$ plateau, respectively.
In principle, $\Psi$ could be evaluated at
different magnitudes, and we choose the H$_2$--convection peak and the
H$^-$ plateau because these features are universal, being determined by stellar
structure.
Thus $\zeta_{\rm K}\approx\Psi(M_{\rm K,1}) / \Psi(M_{\rm K,2})$,
where $\Psi(M_{\rm, K})\,dM_{\rm K}$ is the number of systems in the
magnitude interval $M_{\rm K}$ to $M_{\rm K}+dM_{\rm K}$, $M_{\rm K,1}\approx7$
and $M_{\rm K,2}\approx4.4$.

{}From the bottom panel of Fig.~10 we
note that $\zeta_{\rm K}=8.5/2=4.3$ for the model system  luminosity function
of the Galactic field, whereas $\zeta_{\rm K}=3.5/1.7=2.1$ for the model system
luminosity
function in the central 2~pc sphere of our dynamically highly evolved cluster.
Thus, $\zeta$ is a measure of the state of dynamical evolution of any cluster
provided
the stellar mass function and initial proportion of binaries is universal.

\nobreak\vskip 10pt\nobreak
\noindent{\bf 6.2 Comparison with the Hyades
and Pleiades Clusters}
\nobreak\vskip 10pt\nobreak
\noindent
The stellar populations in open clusters lack the considerable
disadvantage of cosmic scatter inherrent to studies of the luminosity function
in the Galactic disk. Stars in Galactic clusters all have approximately the
same age,  metallicity and distance. The effects of unresolved binary stars
are, however, also severe (Kroupa \& Tout 1992), and the bottom panel of
Fig.~10
shows that cluster luminosity functions suffer from mass segregation and
evaporation of stars. Thus,
when contemplating the universality of the stellar mass function we need to
keep the shortcomings of each sample in mind.

We consider recent determinations of the luminosity function in two open
clusters. Hambly, Hawkins \& Jameson (1991) and Reid (1993) measure the
luminosity functions in the Pleiades and Hyades clusters, respectively.
Leggett,
Harris \& Dahn (1994) obtain a luminosity function for the Hyades for $M_{\rm
V}\ge11$ which is consistent with the data of Reid (1993).
The luminosity function of the Pleiades cluster is shown in
apparent I-band magnitudes as solid circles in the top panel of Fig.~11. The
Hyades luminosity function, converted to absolute V-band magnitudes by Reid
(1993), is shown as solid circles in the bottom panel of Fig.~11.

\nobreak\vskip 10pt\nobreak
\noindent{\bf 6.2.1 Mass segregation and distance estimation}
\nobreak\vskip 10pt\nobreak
\noindent
In both panels of Fig.~11 we plot our model for the Galactic field system
luminsity
function $k\,\Psi_{\rm mod,sys}(t=1\,{\rm Gyr})$, our initial
system luminosity function $k\,\Psi_{\rm mod,sys}(t=0)$,
and also
the fully resolved, i.e. single star, luminosity function, $k\,\Psi_{\rm mod,
sing}$. The constant $k$ we determine by scaling
$\Psi_{\rm mod,sys}(t=1\,{\rm Gyr})$ to the data at $m_{\rm
I}\approx13$ in the top panel, and at $M_{\rm V}\approx10$ in the bottom panel.
In the top panel we assume a distance modulus $m-M=5.5$.

The observed luminosity function for the Pleiades cluster has the shape we
expect, apart from a small overabundance of relatively bright systems at
$m_{\rm I}\approx12.5$. However, its peak lies at $m_{\rm I}=15$, whereas in
our model it lies at $m_{\rm I}=14.4$. If we were to model pre-main sequence
brightening then the discrepancy between our model and the observed luminosity
function would be larger still. Comparing their luminosity function with the
Galactic field photometric luminosity function, Hambly et al. (1991) conclude
that the stellar mass function in the Pleiades cluster must be similar to the
mass function of stars in the field. Assuming the same stellar mass function
for the
Pleiades as for the Galactic field, four reasons might be responsible for the
discrepancy in the location of the peak: (i) Our mass--$M_{\rm V}$ relation is
wrong. However, comparison of the model luminosity functions with the Galactic
field luminosity function in fig.~1 of Kroupa (1995c) suggests that the peak
in our model
cannot be wrong by more that $\Delta M_{\rm V}\approx0.3\,$mag, i.e. $\Delta
m_{\rm I}\approx0.2\,$mag.
(ii) The $M_{\rm V},V-I$ relation derived by Stobie et al.
(1989) using trigonometric parallax data might be systematically wrong.
However, that their relation is a reasonable approximation is easily
demonstrated by plotting
it together with the data published by Monet et al. (1992). To check whether we
have made a mistake when transforming mass to $M_{\rm V}$ to $M_{\rm I}$ to
$m_{\rm I}$ we also explicitly transform the luminosity function of Stobie et
al. (1989) to $m_{\rm I}$ and show it in the top panel of Fig.~11
as crosses, after scaling as above. The same discrepancy is evident.
Leggett et al. (1994) point out that, at $V-I\approx3$, the $M_{\rm V},V-I$
relation appears to steepen. This would alleviate some of the discrepancy found
here. We also note that Kroupa et al. (1993) show that the $M_{\rm V},V-I$
relation derived by Stobie et al. (1989) has to be corrected for systematic
bias owing to cosmic scatter and unresolved binary systems. Even the most
extreme corrected relation (table~6 in Kroupa et al. 1993), however, does not
change $M_{\rm I}$ at $M_{\rm V}=12$ by more than 0.16~mag.
(iii) The photometric calibration of the photographic data might be inaccurate.
The
photographic data were obtained in the R-band, and in their fig.~8, Hambly et
al. (1991) compare their luminosity function with the
Galactic field photographic luminosity function observed in the R-band by
Hawkins \& Bessell (1988). Both data sets agree, suggesting both make the same
systematic error, or point (ii) above.
(iv) Following the suggestion by Kroupa \& Tout (1992) that
the `H$_2$--convection peak' in the stellar luminosity function may be used as
a distance indicator we consider the possibility that
the distance modulus of the Pleiades might be closer to $m-M=6$
than to~5.5. Although the most often quoted value is~5.5 there is a recent
measurement suggesting $m-M=5.9\pm0.26$ (Gatewood et al. 1990, see also
Giannuzzi 1995). In the top panel
of Fig.~11 we plot, as the short-long dashed curve, the model Galactic field
system luminosity function, $k\,\Psi_{\rm mod,sys}(t=1\,{\rm Gyr})$, assuming a
distance modulus of $m-M=6$, and obtain a much improved representation of the
data.

The observed luminosity function for the Hyades cluster has a peak at
the correct location. The overabundance of bright stars is very
apparent, and is interpreted by Reid (1993) to be due to mass segregation and
stellar evaporation. In his elegant paper, Eggen (1993) also shows that the
luminosity function for Hyades stars is depleted at the faint end when compared
to the observed field star luminosity function.

In what follows we adopt a distance modulus of $m-M=6$ for the Pleiades.

\nobreak\vskip 10pt\nobreak
\noindent{\bf 6.2.2 Dynamical age estimation via luminosity function fitting}
\nobreak\vskip 10pt\nobreak
\noindent
We now focus our attention on the system luminosity function
within a central sphere with radius 5~pc in our
dominant mode cluster.
This volume corresponds approximately to the survey volumes of both the Hambly
et al. (1991) and Reid (1993) samples. We plot in Fig.~12 the model
system luminosity functions at times $t=87, 260$ and~476~Myr (they are
tabulated in Table~A-3).

As our cluster evolves we observe in Fig.~12 a
drop in the number of stars (compare to top panel of Fig.~4), and an
increasing deficiency in faint stars. From
Fig.~12 we see
that the Hyades cluster is best represented by a
dynamically advanced model. For the Pleiades we obtain a dynamical age between
90~and 260~Myrs (i.e. between~8 and~24 initial relaxation times), and for the
Hyades about 500~Myrs (i.e. about 44~initial relaxation times).

The fainter location of the peak in the observational Hyades data can
be accounted for by the higher metallicity of the Hyades stars, which
have [Fe/H]$\approx0.15$ (VanDenBerg \& Poll 1989, see also fig.~5 in
Kroupa et al. 1993). The Pleiades cluster has [Fe/H]=0.03 (Cayrel de
Strobel 1990), which cannot account for the residual $\approx0.4\,$mag
fainter location of the peak in the luminosity function in Fig.~12.

For each of the models shown in Fig.~12 we parametrise the depletion of
low-mass stars by evaluating $\zeta$ (equation~6:
$\zeta_{\rm V} = {\Psi_{\rm max} \over \Psi(M_{\rm V}=7)}$ and
$\zeta_{\rm I} = {\Psi_{\rm max} \over \Psi(M_{\rm I}=5.5)}$, where $\Psi_{\rm
max}$ is the maximum of the luminosity function).
The results are plotted in Fig.~13.
For the Hyades and Pleiades data (Fig.~11) we evaluate

$$\zeta_{\rm V}^{\rm Hy} = {\Psi(M_{\rm V}=11.75)\pm\delta\Psi \over
\Psi(M_{\rm V}=6.75)\pm\delta\Psi} = 2.24\pm0.20, $$

\noindent and

$$\zeta_{\rm I}^{\rm Pl} = {\Psi(m_{\rm I}=15.0)\pm\delta\Psi \over
\Psi(m_{\rm I}=11.45)\pm\delta\Psi} = 4.42\pm0.35, $$

\noindent where $\delta\Psi$ is the Poisson uncertainty at the respective
magnitude. The nuclear ages  of the Hyades and
Pleiades Clusters based on isochrone fitting are,
respectively, $655\pm55$~Myrs and $100\pm30$~Myrs (Cayrel de Strobel 1990). The
data are compared with our $\zeta(t)$ model in Fig.~13.

\nobreak\vskip 10pt\nobreak
\noindent{\bf 6.2.3 Discussion}
\nobreak\vskip 10pt\nobreak
\noindent
Concerning the comparison of our models with the observed Hyades and
Pleiades luminosity
functions we must, apart from not modelling the higher metallicity of the
Hyades and high mass stars, keep in mind the following caveats:
(i) The
dynamical evolution of real clusters is affected by perturbations from passing
molecular clouds.
(ii) The published Hyades (and to a lesser degree Pleiades) luminosity
functions
may be incomplete or contaminated by Galactic field stars (see discussion in
Reid 1993).
(iii) The Hyades cluster has
a distance of about 46~pc (VanDenBerg \& Poll 1989), so that some binary
systems
are probably resolved (a binary system with a mass of $0.64\,M_\odot$ and
log$_{10}P>5.2$ has a semi major axis $a>48$~AU, i.e. larger than 1~arc sec).
(iv) The real clusters are likely
to have had an initial $R_{0.5}$ different to that assumed here (0.85~pc). This
has little affect on $n(t)$ and ${\overline m}(t)$ for relaxed clusters,
but determines
$f_{\rm tot}(t)$, and thus influences the shape of the luminosity function.

We thus
treat the dynamical ages suggested in Fig.~12 with reservation,
although our dynamical dating provides encouraging age estimation (Fig.~13).
Extension of the model
$\zeta(t)$ to $t>500$~Myrs will require simulations with $N_{\rm bin}>200$, but
Fig.~13 suggests that $\zeta(t)$ may continue evolving linearly with the same
slope. Using an alternative approach, Buchholz \& Schmidt-Kaler (1980) suggest
that the radial mass distribution as a function of time can be a reliable age
estimator for open clusters.

The excellent agreement of our model with the observed Hyades luminosity
function (Fig.~12) suggests that the proportion of Hyades systems in the
central 2~pc sphere that are binary stars may be 65~per cent (Fig.~4) which is
larger than in the Galactic field ($47\pm5$~per cent).
This would be consistent with the conclusion by Kroupa \& Tout (1992) that a
large
proportion of binaries may reside in the Praesepe Cluster, which is somewhat
older than the Hyades Cluster.
Similarly, we expect that the
total proportion of binaries in the central 2~pc sphere of the Pleiades Cluster
is probably close to 60~per cent (Fig.~4) and may still be decreasing.

Our models do not suggest that the initial dynamical properties
of the stellar systems born in the Hyades and Pleiades Clusters
were significantly different from the Galactic field birth population.
It would thus appear that
we can, in principle, estimate the initial number
of stars that formed in the two clusters (see
Section~4). In doing this, we have to keep the above caveats in mind.

To verify our procedure we first of all consider our models for which we have
complete data. Our $t=476$~Myrs system luminosity function contains 54~systems
with $M_{\rm V}\ge5$ within the central 5~pc sphere (Table~A-3,
or Fig.~12). From Fig.~4 we estimate
$f_{\rm tot}\approx0.65$, so that the number of stars in the central 5~pc
sphere is $N_{5 {\rm pc}}=89$. Comparison with Fig.~4 allows us to estimate the
spatial configuration correction factor to be $s=N_{2 {\rm pc}}(476\,{\rm
Myrs})/N_{5
{\rm pc}}(476\,{\rm Myrs})\approx 40/89 = 0.45$. We can now map
our
`observed' 89 stars in the central 5~pc sphere to the log$_{10}n(t)$ curve in
Fig.~4, and apply equation~1 to solve for the number of stars in the cluster at
birth. The result is $0.45\times89\times{\rm exp}[476/230] = 317$ stars. This
compares favourably with our initial 400~stars.

In the Hyades Cluster Reid (1993) counts about 210~systems with
$M_{\rm V}\ge5$ in the central 5~pc sphere. Following the above method, with
$s=0.45, f_{\rm tot}=0.7$ and adopting the nuclear age, we estimate
$0.45\times357\times{\rm exp}[655/230]=2771$ stars at birth.
Since the mass of our cluster is $128\,M_\odot$, we estimate
the mass of the Hyades Cluster to have been roughly $900\,M_\odot$ at
birth. Stars more massive than $1.1\,M_\odot$ contribute about 30--40~per
cent to the total mass, so that the birth mass of the stellar component of the
Hyades Cluster may have been about $1300\,M_\odot$, in good agreement with
Reid's (1993) estimate. Studying the
distribution of white dwarfs and estimating their loss from the cluster,
Weidemann et al. (1992) estimate the number of stars to have been about
3000--4000 at birth, again in reasonable agreement with our estimate.

Similarly, Hambly et al. (1991) count about 600 systems with $M_{\rm
I}\ge4.5$ in the central 5~pc sphere of the Pleiades Cluster. Assuming
$f_{\rm tot}=0.7$ (Fig.~4), we estimate that the Pleiades may have
contained about $1020\times{\rm exp}[100/230] = 1580$ stars at birth.
Consulting the top panel of Fig.~4 we find that the clusters have not
fully relaxed at $t=100\,$Myrs. We omit $s$ in our present estimate
for $R_{\rm 0.5}<1\,$pc initially (i.e. Hambly et al. have probably
counted virtually all Pleiades members if $R_{0.5}<1\,$pc). For
initially larger $R_{0.5}$, $n(t)$ does not change significantly
during $t<150\,$Myrs.
The total stellar birth mass may thus have been about
$700\,M_\odot$. Limber (1962a) estimates a total present stellar mass
of about $800\,M_\odot$, and van Leeuwen (1980) derives a present
value of about $2000\,M_\odot$. Both estimates assume virial
equilibrium, the latter implying a significantly larger birth mass
than our estimate. Further numerical simulations will be needed to
clarify this issue.

\nobreak\vskip 10pt\nobreak
\noindent{\bf 6.3 The Trapezium Cluster}
\nobreak\vskip 10pt\nobreak
\noindent
The Trapezium Cluster has been extensively observed and dated (Zinnecker,
McCaughrean \& Wilking 1993, Prosser et al. 1994). Prosser et al. find for the
central sphere with a radius of 0.25~pc that the
proportion of binaries with projected separations in the range of approximately
44--440~AU is $f\approx0.11$ which is similar to the Galactic field.
Assuming a system mass of $1.3\,M_\odot$ this separation range corresponds to
$5<{\rm log}_{10}P<6.5$. Their result would appear to be a
lower limit (i.e. $f_{\rm tot}(t)>0.11$, with $t\approx1$~Myr) because they
include all apparently single
stars down to the flux limit. These may have fainter undetected companions.

The present central number density is log$_{10}n_{\rm c}\approx4.5$ so that the
roughly 1~Myrs old Trapezium
Cluster may be compared with our $R_{0.5}=0.08$ and 0.25~pc models (Table~1).
For these clusters we see from fig.~3 in K1 and Fig.~3 here that most
destruction of orbits occurs within
the first few Myrs ($R_{0.5}=0.08$~pc) and within the first few tens of Myrs
($R_{0.5}=0.25$~pc), and from fig.~5 in K1 we deduce that the distribution of
periods
must be similar to the G~dwarf distribution for log$_{10}P<6$ after a few Myrs
and about 20~Myrs, respectively. We remember that fig.~5 needs to be
modified at log$_{10}P<3$
as suggested in sections~5 in K1 and~2.2 in K2. From fig.~5c in K1 we expect
that the Trapezium
Cluster contains no binaries with approximately log$_{10}P>7$ and from fig.~4b
in K1 we expect the mass ratio distribution to be depleted significantly at
small
values. More detailed comparisons and predictions will be possible when more
realistic initial conditions such as the presence of massive stars and a
changing background potential are included.

The binary population of the
Trapezium Cluster is undergoing rapid stimulated evolution (fig.~3 in K1),
given
that it is between 2~and 10~crossing times old (Table~1), and will prove
an
interesting laboratory for the study of the interplay of stimulated evolution
and eigenevolution (cf. to discussion of the $e-{\rm log}_{10}P$ diagram in
section~3.1 in K2).

Comparison of our model luminosity function with the observed Trapezium Cluster
luminosity function by Prosser et al. (1994) is not attempted here owing to the
very difficult and uncertain treatment of pre-main sequence luminosity
evolution (Kroupa, Gilmore \& Tout 1992).
We refer the reader to Zinnecker et al. (1993), who show that the stellar
luminosity function for a sample of stars younger than 2~Myrs can be
significantly distorted because deuterium burning delays contraction.

\vfill\eject

\bigskip
\bigbreak
\noindent{\bf 7 CONCLUSIONS}
\nobreak\vskip 10pt\nobreak
\noindent
Obtaining observational data on clusters of stars is
very important because the same age, metallicity and distance of the stars ease
analysis. A cluster of stars is a fossil of one star-formation event.
We need to study as many of these as is possible in order to
find out, if there is variation of the spectrum of masses formed, or of the
proportion of binaries, and of their distribution of periods. The stellar
population in one cluster samples the birth distribution of dynamical
properties of stellar systems ${\cal D}$ (section~4 in K1).
In this paper we
identify generic features of the evolution of the stellar number density, mean
stellar mass, stellar luminosity function and binary proportion in
Galactic clusters with the special aim of addressing observable properties.

We assume that all stars form in binary systems with component
masses paired at random from the KTG(1.3) mass function.
The initial period distribution is flat in the range
$3\le{\rm log}_{10}P\le 7.5$, $P$ in days. For the purposes of the present
study, we assume that 200 binaries initially populate stellar clusters
with half mass radii 0.08~pc$\,\le R_{0.5}\le2.5$~pc that
are initially in virial equillibrium. We
consider stellar masses in the range $0.1\,M_\odot\le m\le1.1\,M_\odot$
to avoid complications concerning stellar evolution. We also distribute
400 single stars for a comparison with the binary star clusters. Using
direct N-body integration we follow the evolution of these stellar systems
until they disperse, repeating the experiment five times for the binary star
clusters and three times for the single star clusters.
 We also perform 20~simulations of the
dominant mode $(N_{\rm bin},R_{0.5})=(200,0.85\,{\rm pc})$ cluster, in which
the binary stars have a rising period distribution with increasing log$_{10}P
\ge 0.48$. The parameters for these clusters are listed in Table~1.

We find that the number density evolution in the central 2~pc sphere of both
the binary and single star clusters
is indistinguishable (Fig.~1) and has an exponential decay time $\tau_{\rm
e}\approx230$~Myrs. Both decay to less than
0.1~stars~pc$^{-3}$
within about $700$~Myr. Cluster lifetime and evolution of number density
within the central 2~pc sphere are invariant to changes of initial $R_{0.5}$ if
$0.08\le R_{0.5}\le2.53$~pc (Section~3).
After the initial rapid ionisation of the less bound binary systems the
proportion of binary systems rises slowly in the central region of the
clusters (Figs.~3 and~4). However, even for our most compact cluster
($R_{0.5}=0.08$~pc) the binary proportion in the central region does not
decrease below about 30~per cent. The central proportion of binary systems in a
cluster
is a sensitive function of the initial concentration of the cluster, whereas
segregation of mean stellar mass is a sensitive function of the dynamical age
of the cluster (Figs.~2 and~4). Evolution of the number density and mean
stellar mass, and scaling to other initial cluster parameters, are explored in
Section~4.

The {\it kinematical signature of star
formation} reflects the initial configuration (Fig.~5): the clusters consisting
initially of
100~per~cent
primordial binaries lead to a different distribution of centre of mass kinetic
energies than the clusters that initially have no primordial binaries. The
former have a distinct high kinetic energy tail which is a
function of the initial cluster configuration. The high velocity tail
ought to be apparent in
the distribution of young stars in the vicinity of star forming
regions.

If the majority of stars form in aggregates that are dynamically equivalent to
the
dominant mode cluster then only a few per cent of all stars are ejected from
the cluster with a large enough velocity to escape the molecular cloud.
Roughly 90~per cent of all stars have a velocity smaller than about 5~km
sec$^{-1}$ and remain
trapped in the vicinity of their parent giant molecular cloud, until it
disperses after a few
$10^7$ years. This population of stars may appear as a distributed population
of pre-main sequence stars either while star formation continues in the cloud,
or after it has ceased. We expect about 50~per cent of young stellar
systems in the apparently distributed population to be binaries.
In this light, it seems possible that the L1630
molecular cloud in the Orion complex is void of a distributed population
of young stars because star formation has only just begun in the four
locations,
where Lada \& Lada (1991) report embedded clusters which are similar to our
dominant mode cluster (K1). The L1641 molecular cloud in the southern region of
Orion A may have been forming stars over a time span of about 7~Myrs, which
may explain the significant distributed population of young stars with an age
of 5-7~Myrs which Strom et al. (1993) detect. The stellar dynamical properties
of this population will prove very useful in discriminating its birth dynamical
structures.

A halo of young stars around a
molecular cloud may be expected because even stars with escape velocities
(implying a close encounter with a binary system and probable loss of
circumstellar
material thus becoming a weak-lined T-Tauri star, i.e. wTTS) may remain trapped
by the equipotential surface of the molecular cloud and Galaxy. Only
about 15~per cent of these systems are binaries with the proportion
of binaries decreasing with increasing ejection velocity (Fig.~9).  The
discovery of
wTTS distributed over the entire region of the Orion molecular cloud complex by
Sterzik et al. (1995) is thus of particular interest.

If most stars are born in embedded clusters similar to our dominant
mode cluster then we require a birth rate of roughly 15~clusters~kpc$^{-2}$
Myr$^{-1}$ (Section 3.1).

The stellar luminosity function in the central region
of the dominant mode cluster flattens with time (i.e. an overabundance of
bright systems develops) although the `H$_2$--convection peak' at
$M_{\rm V}\approx12, M_{\rm K}\approx6.5$
remains. Assuming a universal initial mass function we thus expect the
luminosity
function in Galactic clusters to differ from that in the Galactic field
(Figs.~10 and~11). Recently published data on the Hyades luminosity
function (Reid 1993) are consistent with a dynamically evolved cluster
(Fig.~12). The location of
the peak in the Pleiades luminosity function data (Hambly et al. 1991) suggests
a distance
modulus of $m-M=6$ rather than 5.5 (top panel of Fig.~12).

Parametrising stellar mass dependent evaporation from the central
region of the dominant mode
cluster by the ratio $\zeta$ (equation~6) of the luminosity function at the
`H$_2$--convection peak'
and at the `H$^-$ plateau', we find good
agreement with $\zeta(t)$ obtained from the Hyades and Pleiades luminosity
functions (Fig.~13). This result, together with our results on the evolution of
the stellar number density, mean stellar mass and binary star proportion,
suggests: (i) that our initial
assumptions about stellar mass function and binary stars are consistent with
the data, and (ii) for each cluster a unique evolutionary history
probably exists which may be found by simple scaling to pre-computed histories
(Section 4).
Thus, using Fig.~4 we may expect that about 60--70~per cent of
all systems in the central 2~pc sphere are binaries in the Pleiades and Hyades
Clusters. A rough estimate of the initial number of stars in the
Hyades and Pleiades Clusters suggests birth masses of roughly
$1200\,M_\odot$ and $700\,M_\odot$, respectively. The former value is
consistent with other estimates (Weidemann et al. 1992, Reid 1993), but the
latter value is significantly smaller than the estimate by van
Leeuwen (1980).

The simulation of a cluster with high initial stellar number density
($R_{0.5}<0.25$~pc) implies
that the period distribution in the range log$_{10}P<6$ is similar to the main
sequence period distribution after a few initial relaxation times, which is
consistent with
observations of the Trapezium Cluster (Section~6.3). The binary population in
this cluster must be undergoing significant stimulated evolution.

More detailed comparisons with observational data (e.g. colur--magnitude
diagrams, spatial distribution of bright and faint stars, spatial distribution
of velocities), and modeling of individual Galactic clusters, as well as
embedded clusters, and including massive
stars, a varying background potential during the first 10~Myrs and non-virial
equilibrium initial conditions, will mature our present ideas and results. By
adopting the nuclear ages of clusters and comparing with realistic N-body
models, we can hope to identify the physics responsible for shaping the
cluster.

\bigskip
\bigbreak
\par\noindent{\bf ACKNOWLEDGMENTS}
\nobreak
\nobreak
\noindent I am very grateful to Sverre Aarseth for
kindly helping with and distributing nbody5, and to P. J. T. Leonard for
valuable comments. I also thank H. Reffert for useful discussions, and am
grateful to Peter Schwekendiek for his rapid acquaintance with and
maintenance of the new computing facilities installed at ARI towards the end of
1993.

\bigskip
\noindent{\bf REFERENCES}
\nobreak
\bigskip
\nex Aarseth, S. J., 1988a, Boletin de la Academia Nacional de Ciencias, 58,
     177
\nex Aarseth, S. J., 1988b, Boletin de la Academia Nacional de Ciencias, 58,
     189
\nex Aarseth, S. J., 1994, Direct Methods for N-Body Simulations. In:
      Contopoulos, G., Spyrou, N. K., Vlahos, L. (eds.), Galactic Dynamics and
      N-Body simulations, Springer, Berlin, p.277
\nex Aarseth, S. J., Hills, J. G., 1972, A\&A 21, 255
\nex Battinelli, P., Capuzzo-Dolcetta, R., 1991, MNRAS 249, 76
\nex Battinelli, P., Brandimarti, A., Capuzzo-Dolcetta, R., 1994, A\&AS 104,
     379
\nex Blitz, L., 1993, Giant Molecular Clouds. In:
     Levy, E. H., Lunine,
     J. I. (eds), Protostars and Planets III, Univ. of Arizona Press, Tucson,
     p. 125
\nex  Buchholz, M., Schmidt-Kaler, Th., 1980, Dynamical Age Estimation of Open
      Clusters. In: Hesser, J. E. (ed.), IAU Symp.85, Star Clusters, Reidel,
      Dordrecht, p.221
\nex Cayrel de Strobel, G., 1990, Mem.S.A.It. 61, 613
\nex Eggen, O. J., 1993, AJ 106, 1885
\nex Gatewood, G., Castelaz, M., Han, I., Persinger, T., Stein, J., Stephenson,
     B., Tangren, W., 1990, ApJ, 364, 114
\nex Giannuzzi, M. A., 1995, A\&A 293, 360
\nex Griffin, R. F., Gunn, J. E., Zimmerman, B. A., Griffin, R. E. M., 1988,
     AJ 96, 172
\nex Hambly, N. C., Hawkins, M. R. S., Jameson, R. F., 1991, MNRAS, 253, 1
\nex Hawkins, M. R. S., Bessell, M. S., 1988, MNRAS, 234, 177
\nex Heggie, D. C., 1975, MNRAS 173, 729
\nex Heggie, D. C., Aarseth, S. J., 1992, MNRAS 257, 513
\nex Hills, J. G., 1975, AJ 80, 809
\nex Hut, P., 1985, Binary Formation and Interactions with Field Stars. In:
     Goodman, J., Hut,
     P. (eds.), Proc. IAU Symp 113, Dynamics of Star Clusters, Reidel,
     Dordrecht, p.231
\nex Hut, P., McMillan, S., Goodman, J., et al., 1992, PASP 104, 981
\nex Hut, P., McMillan, S., Romani, R. W., 1992, ApJ 389, 527
\nex Kroupa, P., 1995a, Inverse Dynamical Population Synthesis and Star
     Formation, in preparation (K1)
\nex Kroupa, P., 1995b, The Dynamical Properties of Stellar Systems in the
     Galactic Disc, in preparation (K2)
\nex Kroupa, P., 1995c, Unification of the Nearby and Photometric Stellar
     Luminosity Functions, Nov.~1 issue of ApJ
\nex Kroupa, P., Tout, C. A., 1992, MNRAS 259, 223
\nex Kroupa, P., Gilmore, G., Tout, C. A., 1992, AJ 103, 1602
\nex Kroupa, P. Tout, C. A., Gilmore, G., 1990, MNRAS 244, 76
\nex Kroupa, P., Tout, C. A., Gilmore, G., 1993, MNRAS 262, 545
\nex Lada, C. J., Lada, E. A., 1991, The Nature, Origin and Evolution of
     Embedded
     Star Clusters. In: Janes, K. (ed.), The Formation and Evolution of Star
     Clusters, PASP Conf. Series, Vol.13, San Francsico, p.3
\nex Lada, C. J., Margulis, M., Dearborn, D., 1984, ApJ 285, 141
\nex Leggett, S. K., Harris, H. C., Dahn, C. C., 1994, AJ 108, 944
\nex Leonard, P. J. T., Duncan, M. J., 1990, AJ 99, 608
\nex Leonard, P. J. T., Linnell, A. P., 1992, AJ, 103, 1928
\nex Limber, D. N., 1962a, ApJ 135, 16
\nex Limber, D. N., 1962b, ApJ 135, 41
\nex Mathieu, R. D., 1985, The Structure and Internal Kinematics of Open
     Cluster.
     In: Goodman, J., Hut,
     P. (eds.), Proc. IAU Symp 113, Dynamics of Star Clusters, Reidel,
     Dordrecht, p. 427
\nex Mathieu, R. D., 1986, Highlights of Astronomy 7, 481
\nex McMillan, S., Hut, P., 1994, ApJ, 427, 793
\nex Mermilliod, J.-C., Rosvick, J. M., Duquennoy, A., Mayor, M., 1992, A\&A
     265, 513
\nex Monet, D.G., Dahn, C.C., Vrba, F.J., et al., 1992, AJ 103, 638
\nex Pinto, F., 1987, PASP 99, 1161
\nex Prosser, C. F., Stauffer, J. R., Hartmann, L., Soderblom, D. R., Jones, B.
     F., Werner, M. W., McCaughrean, M. J., 1994, ApJ 421, 517
\nex Reid, N., 1993, MNRAS, 265, 785
\nex Steele, I. A., Jameson, R. F., 1995, MNRAS 272, 630
\nex Sterzik, M. F., Alcala, J. M., Neuh{\"a}user, R., Schmitt, J. H. M. M.,
      1995, A\&A 297, 418
\nex Stobie, R. S., Ishida, K., Peacock, J. A., 1989, MNRAS, 238, 709
\nex Strom, K. M., Strom, S. E., Merrill, K. M., 1993, ApJ 412, 233
\nex Terlevich, E., 1987, MNRAS 224, 193
\nex Theuns, T., 1992, A\&A 259, 503
\nex van Leeuwen, F., 1980, Mass and Luminosity Function of the Pleiades,
     In: Hesser, J. E. (ed.), IAU Symp.85, Star Clusters, Reidel,
     Dordrecht, p.157
\nex VandenBerg, D. A., Poll, H. E., 1989, AJ, 98, 1451
\nex Verschueren, W., David, M., 1989, A\&A 219, 105
\nex Weidemann, V., Jordan, S., Iben, I., Casertano, S., 1992, AJ 104, 1876
\nex Wielen, R., 1971, A\&A 13, 309
\nex Wielen, R., 1985, Dynamics of Open Star Clusters. In: Goodman, J., Hut,
     P. (eds.), Proc. IAU Symp 113, Dynamics of Star Clusters, Reidel,
     Dordrecht, p.449
\nex Wielen, R., 1988, Dissolution of Star Clusters in Galaxies. In: Grindlay,
     J. E., Davis Philip, A. G. (eds.), Globular Cluster Systems in Galaxies,
     Proc. IAU Symp 126,  Kluwer, Dordrecht, p.393
\nex Zinnecker, H., McCaughrean, M. J., Wilking, B. A., 1993, The Initial
     Stellar Population. In:
     Levy, E. H., Lunine,
     J. I. (eds), Protostars and Planets III, Univ. of Arizona Press, Tucson,
     p. 429

\vfill\eject

\bigbreak
\vskip 3mm
\bigbreak

\hang{ {\bf Table A-1: The Distribution of Centre of Mass Kinetic Energies
After Cluster Dissolution} (Section~5.1)}

\nobreak
\vskip 1mm
\nobreak
{\hsize 15 cm \settabs 9 \columns

\+ &&&&$R_{0.5}$\cr
\+\cr

\+log$_{10}E_{\rm kin}$ &2.53 &0.77 &0.25 &0.08 &0.25 &0.08  &
log$_{10}E_{\rm kin}$  &0.85\cr

\+$M_\odot$ km$^2$
&pc &pc &pc &pc &pc &pc
&$M_\odot$ km$^2$  &pc \cr
\+sec$^{-2}$  &&&&&&&sec$^{-2}$\cr
\+$N_{\rm bin}=$  &200 &200 &200 &200 &0   &0   & &200\cr
\+$N_{\rm sing}=$ &0   &0   &0   &0   &400 &400 & &0\cr

\+\cr
\+$-4.68$ &0.0  &0.0   &0.0   &0.0   &0.0    &0.0    &$-4.80$  &0.0\cr
\+$-4.32$ &0.0  &0.0   &0.0   &0.0   &0.0    &0.0    &$-4.46$  &0.0\cr
\+$-3.96$ &0.0  &0.0   &0.2   &0.6   &0.3    &0.0    &$-4.13$  &0.0\cr
\+$-3.61$ &0.0  &0.6   &0.0   &0.4   &0.6    &2.0    &$-3.79$  &0.1\cr
\+$-3.25$ &1.0  &1.2   &2.6   &2.0   &4.0    &3.3    &$-3.45$  &0.5\cr
\+$-2.89$ &3.4  &2.8   &5.4   &3.4   &8.3    &8.0    &$-3.11$  &1.9\cr
\+$-2.54$ &9.0  &11.4  &14.0  &13.0  &32.0   &23.0   &$-2.77$  &5.7\cr
\+$-2.18$ &20.0 &26.0  &34.8  &26.6  &63.0   &56.0   &$-2.43$  &15.0\cr
\+$-1.82$ &36.0 &44.4  &50.4  &52.0  &87.6   &88.3   &$-2.10$  &31.1\cr
\+$-1.46$ &50.2 &51.0  &58.0  &54.6  &91.3   &79.3   &$-1.76$  &49.2\cr
\+$-1.11$ &46.2 &50.8  &41.6  &50.8  &56.0   &56.0   &$-1.42$  &56.3\cr
\+$-0.75$ &32.0 &25.4  &24.8  &37.0  &23.0   &31.3   &$-1.08$  &48.7\cr
\+$-0.39$ &10.0 &12.4  &12.4  &15.2  &8.0    &14.6   &$-0.74$  &24.7\cr
\+$-0.04$ &4.2  &8.0   &9.4   &14.4  &5.0    &9.6    &$-0.40$  &10.8\cr
\+$+0.32$ &2.2  &4.0   &9.6   &10.4  &5.0    &6.3    &$-0.07$  &4.9 \cr
\+$+0.68$ &1.8  &4.8   &9.0   &9.0   &3.0    &4.6    &$+0.27$  &4.8 \cr
\+$+1.04$ &0.2  &2.6   &5.2   &8.0   &2.6    &2.6    &$+0.61$  &3.2 \cr
\+$+1.39$ &0.2  &1.2   &3.4   &7.8   &1.6    &2.3    &$+0.95$  &2.8 \cr
\+$+1.75$ &0.0  &0.6   &2.0   &4.2   &0.0    &2.3    &$+1.29$  &2.1 \cr
\+$+2.11$ &0.0  &0.6   &1.8   &3.2   &0.3    &1.3    &$+1.63$  &1.0 \cr
\+$+2.46$ &0.0  &0.0   &0.6   &1.4   &0.0    &0.0    &$+1.96$  &0.5 \cr
\+$+2.82$ &0.0  &0.0   &0.0   &0.6   &0.0    &0.0    &$+2.30$  &0.4 \cr
\+        &     &      &      &      &       &       &$+2.64$  &0.2 \cr
\+        &     &      &      &      &       &       &$+2.98$  &0.1 \cr
\+        &     &      &      &      &       &       &$+3.32$  &0.0 \cr

}
\bigbreak\vskip 3mm

\hang{The $R_{0.5}=0.85$~pc data are binned somewhat differently for historical
reasons}

\vfill\eject

\bigbreak
\vskip 3mm
\bigbreak

\hang{ {\bf Table~A-2: Velocity Distribution After Disintegration of the
Dominant Mode Cluster} (Section~5.2)}

\nobreak
\vskip 1mm
\nobreak
{\hsize 15 cm \settabs 6 \columns

\+&~~log$_{10}v$     &~~$f_v$ &~~$\delta f$ &~$<m>_v$ &$N_{{\rm
sing},v}$
&$N_{{\rm bin},v}$\cr
\+&km~sec$^{-1}$ & & &~~$M_\odot$\cr
\+\cr
\+&$-1.775$ &0.0000&  0.0000 &0.000 &  0 &  0 \cr
\+&$-1.625$ &0.0004&  0.0003 &0.799 &  0 &  2 \cr
\+&$-1.475$ &0.0004&  0.0003 &0.278 &  1 &  1 \cr
\+&$-1.325$ &0.0015&  0.0005 &0.555 &  0 &  8 \cr
\+&$-1.175$ &0.0055&  0.0010 &0.688 &  6 & 23 \cr
\+&$-1.025$ &0.0138&  0.0017 &0.645 & 18 & 55 \cr
\+&$-0.875$ &0.0299&  0.0024 &0.617 & 28 &130 \cr
\+&$-0.725$ &0.0658&  0.0036 &0.608 & 95 &253 \cr
\+&$-0.575$ &0.1225&  0.0049 &0.534 &225 &423 \cr
\+&$-0.425$ &0.1641&  0.0057 &0.500 &353 &515 \cr
\+&$-0.275$ &0.1934&  0.0062 &0.469 &505 &518 \cr
\+&$-0.125$ &0.1664&  0.0058 &0.451 &534 &346 \cr
\+&$+0.025$ &0.1043&  0.0046 &0.401 &408 &144 \cr
\+&$+0.175$ &0.0486&  0.0031 &0.439 &211 & 46 \cr
\+&$+0.325$ &0.0170&  0.0018 &0.601 & 69 & 21 \cr
\+&$+0.475$ &0.0198&  0.0020 &0.664 & 84 & 21 \cr
\+&$+0.625$ &0.0136&  0.0016 &0.684 & 50 & 22 \cr
\+&$+0.775$ &0.0108&  0.0015 &0.496 & 47 & 10 \cr
\+&$+0.925$ &0.0076&  0.0012 &0.565 & 34 &  6 \cr
\+&$+1.075$ &0.0059&  0.0011 &0.504 & 26 &  5 \cr
\+&$+1.225$ &0.0036&  0.0008 &0.736 & 14 &  5 \cr
\+&$+1.375$ &0.0026&  0.0007 &0.293 & 14 &  0 \cr
\+&$+1.525$ &0.0015&  0.0005 &0.740 &  7 &  1 \cr
\+&$+1.675$ &0.0009&  0.0004 &0.817 &  5 &  0 \cr
\+&$+1.825$ &0.0002&  0.0002 &0.184 &  1 &  0 \cr
\+&$+1.975$ &0.0000&  0.0000 &0.000 &  0 &  0 \cr
}
\bigbreak\vskip 3mm

\hang{
The data are plotted in Fig.~9 and are a mean of 20~simulations. Column~1
contains the centre of each velocity bin;
Column~2 the proportion of systems in each velocity bin; Column~3 the Poisson
uncertainty; Column~4 the mean mass per bin; Columns~5 and~6, respectively,
list the number of single stars and binary systems per bin.

\vskip 2mm

$<m>_v = {M_v\over N_{{\rm bin},v}+N_{{\rm sing},v}}$, where $M_v$
is the total mass in velocity bin $v$.
}

\vfill\eject

\bigbreak
\vskip 3mm
\bigbreak

\hang{ {\bf Table~A-3: The System Luminosity Function in the Central
5~pc Sphere of the Dominant Mode Cluster} (Section~6.2.2)}

\nobreak
\vskip 1mm
\nobreak
{\hsize 15 cm \settabs 7 \columns

\+& $M_{\rm V}$ &$\Psi_{\rm mod,sys}$ &$\delta \Psi$
&$\Psi_{\rm mod,sys}$ &$\delta \Psi$
&$\Psi_{\rm mod,sys}$ &$\delta \Psi$ \cr
\+&  &~~~$t=87\,$Myr  & &~~~$t=260\,$Myr & &~~~$t=476\,$Myr \cr

\+\cr
\+&  0.0 &  0.00&  0.00 &     0.00&   0.00 &     0.00&   0.00 \cr
\+&  0.5 &  0.00&  0.00 &     0.00&   0.00 &     0.00&   0.00 \cr
\+&  1.0 &  0.05&  0.05 &     0.00&   0.00 &     0.00&   0.00 \cr
\+&  1.5 &  0.20&  0.10 &     0.10&   0.07 &     0.05&   0.05 \cr
\+&  2.0 &  0.25&  0.11 &     0.10&   0.07 &     0.05&   0.05 \cr
\+&  2.5 &  0.30&  0.12 &     0.10&   0.07 &     0.10&   0.07 \cr
\+&  3.0 &  0.25&  0.11 &     0.05&   0.05 &     0.05&   0.05 \cr
\+&  3.5 &  0.65&  0.18 &     0.40&   0.14 &     0.20&   0.10 \cr
\+&  4.0 &  0.90&  0.21 &     0.55&   0.17 &     0.35&   0.13 \cr
\+&  4.5 &  2.55&  0.36 &     2.40&   0.35 &     1.10&   0.24 \cr
\+&  5.0 &  4.05&  0.46 &     2.85&   0.38 &     1.80&   0.30 \cr
\+&  5.5 &  2.30&  0.34 &     1.40&   0.27 &     0.85&   0.21 \cr
\+&  6.0 &  4.40&  0.48 &     3.45&   0.42 &     1.90&   0.31 \cr
\+&  6.5 &  5.20&  0.52 &     3.40&   0.42 &     2.45&   0.35 \cr
\+&  7.0 &  5.25&  0.52 &     4.25&   0.47 &     2.50&   0.36 \cr
\+&  7.5 &  3.25&  0.41 &     2.05&   0.32 &     1.30&   0.26 \cr
\+&  8.0 &  2.10&  0.33 &     1.60&   0.29 &     0.85&   0.21 \cr
\+&  8.5 &  7.35&  0.62 &     4.85&   0.50 &     2.85&   0.38 \cr
\+&  9.0 &  6.30&  0.57 &     4.70&   0.49 &     2.45&   0.35 \cr
\+&  9.5 &  8.20&  0.65 &     5.80&   0.55 &     2.70&   0.37 \cr
\+& 10.0 &  9.85&  0.72 &     7.00&   0.60 &     3.80&   0.44 \cr
\+& 10.5 & 17.00&  0.94 &    11.25&   0.76 &     4.90&   0.50 \cr
\+& 11.0 & 18.65&  0.99 &    13.25&   0.83 &     6.20&   0.57 \cr
\+& 11.5 & 22.60&  1.09 &    12.40&   0.80 &     4.75&   0.50 \cr
\+& 12.0 & 22.35&  1.08 &    11.60&   0.78 &     4.75&   0.50 \cr
\+& 12.5 & 13.50&  0.84 &     8.25&   0.65 &     2.85&   0.38 \cr
\+& 13.0 & 10.80&  0.75 &     5.70&   0.54 &     1.45&   0.27 \cr
\+& 13.5 &  9.10&  0.69 &     4.50&   0.48 &     1.30&   0.26 \cr
\+& 14.0 &  6.70&  0.59 &     4.10&   0.46 &     1.75&   0.30 \cr
\+& 14.5 &  5.45&  0.53 &     2.85&   0.38 &     0.80&   0.20 \cr
\+& 15.0 &  5.75&  0.55 &     2.25&   0.34 &     0.80&   0.20 \cr
\+& 15.5 &  5.20&  0.52 &     2.70&   0.37 &     0.75&   0.19 \cr

\vfill\eject

\+& $M_{\rm I}$ &$\Psi_{\rm mod,sys}$ &$\delta \Psi$
&$\Psi_{\rm mod,sys}$ &$\delta \Psi$
&$\Psi_{\rm mod,sys}$ &$\delta \Psi$ \cr
\+&  &~~~$t=87\,$Myr  & &~~~$t=260\,$Myr & &~~~$t=476\,$Myr \cr
\+\cr
\+&  0.5 &  0.00 & 0.00 &    0.00 &  0.00 &     0.00 & 0.00 \cr
\+&  1.0 &  0.15 & 0.08 &    0.05 &  0.05 &     0.00 & 0.00 \cr
\+&  1.5 &  0.20 & 0.10 &    0.15 &  0.08 &     0.10 & 0.07 \cr
\+&  2.0 &  0.35 & 0.13 &    0.10 &  0.07 &     0.05 & 0.05 \cr
\+&  2.5 &  0.35 & 0.13 &    0.05 &  0.05 &     0.10 & 0.07 \cr
\+&  3.0 &  0.65 & 0.18 &    0.45 &  0.15 &     0.20 & 0.10 \cr
\+&  3.5 &  1.65 & 0.29 &    1.25 &  0.25 &     0.65 & 0.18 \cr
\+&  4.0 &  3.60 & 0.43 &    3.15 &  0.40 &     1.75 & 0.30 \cr
\+&  4.5 &  4.15 & 0.46 &    2.35 &  0.35 &     1.50 & 0.28 \cr
\+&  5.0 &  5.75 & 0.55 &    4.40 &  0.48 &     2.45 & 0.35 \cr
\+&  5.5 &  7.30 & 0.61 &    5.35 &  0.53 &     3.40 & 0.42 \cr
\+&  6.0 &  5.25 & 0.52 &    3.65 &  0.43 &     2.20 & 0.34 \cr
\+&  6.5 &  5.55 & 0.54 &    3.75 &  0.44 &     2.55 & 0.36 \cr
\+&  7.0 & 10.55 & 0.74 &    7.65 &  0.63 &     3.95 & 0.45 \cr
\+&  7.5 & 12.25 & 0.80 &    8.95 &  0.68 &     4.35 & 0.47 \cr
\+&  8.0 & 19.30 & 1.00 &   12.65 &  0.81 &     5.90 & 0.55 \cr
\+&  8.5 & 31.25 & 1.28 &   20.05 &  1.02 &     9.15 & 0.69 \cr
\+&  9.0 & 29.65 & 1.24 &   15.80 &  0.91 &     6.30 & 0.57 \cr
\+&  9.5 & 20.85 & 1.04 &   12.20 &  0.80 &     4.00 & 0.45 \cr
\+& 10.0 & 15.60 & 0.90 &    8.40 &  0.66 &     2.60 & 0.36 \cr
\+& 10.5 &  8.45 & 0.66 &    4.80 &  0.50 &     1.60 & 0.29 \cr
\+& 11.0 &  9.00 & 0.68 &    4.45 &  0.48 &     1.60 & 0.29 \cr
\+& 11.5 &  6.95 & 0.60 &    3.10 &  0.40 &     1.05 & 0.23 \cr
\+& 12.0 &  4.15 & 0.46 &    2.65 &  0.37 &     0.55 & 0.17 \cr

}
\bigbreak\vskip 3mm

\hang{
Pre-main sequence stellar evolution is not modelled.
Column~1 lists the absolute magnitudes. Columns~2 and~3 list the
system luminosity function and the standard deviation of the mean (see
appendix~1 in Kroupa 1995c), respectively, at time
$t=87\,$Myr.
The following two pairs of columns contain the system luminosity function at
260~Myr and 476~Myr. The luminosity functions are
averages of 20~simulations.}

\vfill\eject

\centerline{\bf Figure captions}
\smallskip

\noindent {\bf Figure 1.} Evolution of the number density of stars within the
central 2~pc sphere of the binary star clusters
(shown by the different lines) and the two single star clusters
(shown by the open and solid circles) (Section~3.1). The horizontal
dotted line marks the density
below which the cluster is considered completely disintegrated.

\vskip 5mm

\noindent {\bf Figure 2.} Evolution of mass segregation in the four
binary star clusters ({\bf top four panels}) and the two single star clusters
({\bf bottom two panels}), respectively (Section~3.2).
The mean stellar mass within the central
2~pc sphere is plotted as open circles and
the mean stellar mass outside this sphere is plotted as solid circles. The
apparent drop in the mean mass within the central 2~pc sphere after
approximately 600~Myrs results from our avaraging technique in which the mean
mass is computed from the $N_{\rm run}$ simulations (see table~1 in K1) at a
time when some of the simulations have lead to completely dissolved clusters.

\vskip 5mm

\noindent {\bf Figure 3.} Evolution of the overall proportion of binaries
within (open circles, $f_{\rm in}(t)$) and
outside (filled circles, $f_{\rm out}(t)$) the central 2~pc sphere
for the four binary star clusters (Section~3.2).

\vskip 5mm

\noindent {\bf Figure 4.}  The average evolution of
20~simulations of the dominant mode cluster initially with $(N_{\rm
bin},R_{0.5})=(200,0.85\,{\rm pc})$ (Section~4). {\bf Top panel}: The
number density evolution within the central 2~pc sphere
(solid curve) is compared to the number density
evolution of the clusters discussed in Section~3 (dotted curves, Fig.~1).
{\bf Middle panel}: The evolution of ${\overline m}(t)$ within the central 2~pc
sphere (open circles) and outside this sphere (solid circles). The evolution of
${\overline m}(t)$ inside the central 2~pc sphere for the four binary star
clusters (dotted curve: $R_{0.5}=2.53$~pc; short dashed curve:
$R_{0.5}=0.77$~pc; long dashed curve: $R_{0.5}=0.25$~pc; dot dashed curve:
$R_{0.5}=0.08$~pc) and for
the two single star clusters (dot long dashed curves)
shown in Fig.~2 is also plotted here. {\bf Bottom panel}: The evolution of
the overall proportion of binaries within the central 2~pc sphere
(open circles), and outside this sphere (solid circles). The evolution of
$f_{\rm in}(t)$ for the four binary star clusters shown in
Fig.~3 is plotted here using the same symbols as in the middle panel.

\vskip 5mm

\noindent {\bf Figure 5.} The energy distributions (Section~5.1). Units of
energy are $M_\odot$ km$^2$
sec$^{-2}$. {\bf Top panel:} The initial ($t=0$) distribution of centre of mass
kinetic energies is shown by the solid curves for the two binary star
clusters with initial $R_{0.5}=2.53$~pc (left distribution) and~0.08~pc
(right distribution).
The distributions after cluster dissolution are represented by the dotted curve
($R_{0.5}=2.53$~pc), the short dashed curve ($R_{0.5}=0.77$~pc), the long
dashed curve ($R_{0.5}=0.25$~pc) and the dot dashed curve
($R_{0.5}=0.08$~pc). The final distribution of kinetic energies for the single
star clusters is shown by the solid triangles: long dashed curve
($R_{0.5}=0.25$~pc) and dot dashed curve ($R_{0.5}=0.08$~pc). These have been
scaled to the binary star curves at log$_{10}E_{\rm kin}\approx-1.6$.
The binary star clusters with $R_{0.5}=0.08$ and~0.25~pc
have a significantly
larger relative population with log$_{10}E_{\rm kin}>0$ than the single star
clusters.
{\bf Bottom panel:} The distribution of initial (solid curve) and final binding
energy distributions of binaries for the four binary star clusters. The initial
$R_{0.5}$ are represented by the same symbols as in the top panel. The
increased
erosion of the binary star population is apparent as the initial kinetic energy
distribution (solid curves in top panel) shift to higher energies.

\vskip 5mm

\noindent {\bf Figure 6.} The final centre of mass kinetic (thin histogram) and
binary system binding
(thick histogram) energy distributions for the two binary star clusters with
initial $R_{0.5}=0.25$~pc ({\bf top panel}) and $R_{0.5}=0.08$~pc ({\bf bottom
panel}). Units of energy are $M_\odot$ km$^2$ sec$^{-2}$.

\vskip 5mm

\noindent {\bf Figure 7.} Comparison of the $(N_{\rm
bin},R_{0.5})=(200,0.77\,{\rm pc})$ cluster shown in Fig.~5 with the
$(200,0.85\,{\rm pc})$ cluster (Section~5.1). Although the initial distribution
of binary binding energies ({\bf bottom panel}) differs in both cases the final
distribution of centre of mass kinetic energies ({\bf top panel}) is
indistinguishable. The initial kinetic energy and binding energy distributions
are shown by thin solid curves for the $R_{0.5}=0.77$~pc cluster, and as thick
solid curves with small solid dots for the $R_{0.5}=0.85$~pc cluster. The final
kinetic energy and binding energy distributions are shown as the thin dashed
curves for the $R_{0.5}=0.77$~pc cluster, and as thick dashed curves with large
solid dots for the $R_{0.5}=0.85$~pc cluster. Systems have to overcome the
potential of the cluster so that after cluster dissolution we have an increased
number of systems with log$_{10}E_{\rm kin}<-1.4$.

\vskip 5mm

\noindent {\bf Figure 8.} The distributions of centre of mass kinetic energies
are plotted as thin lines and the
distribution of binding energies of the binaries are plotted as thick lines
(Section~5.1). The mean of 20 simulations is shown for the $R_{0.5}=0.85$~pc
cluster. {\bf Top panel}: The initial distributions.
{\bf Middle panel}: The distributions after cluster dissolution.
{\bf Bottom panel}: The difference of the kinetic energy
distributions is shown as the thin line, and the difference of the binding
energy distributions is shown as the thick line.
Negative numbers correspond to a gain.
Units of energy are $M_\odot$~km$^2$/sec$^2$.

\vskip 5mm

\noindent {\bf Figure 9.} The mean distribution of velocities of
20 simulations of the $R_{0.5}=0.85$~pc cluster (Section~5.2).
{\bf Top panel}: The
initial proportion of systems as a function of their centre of mass velocity
is shown as the dashed histogram.
After cluster disintegration the distribution of centre of mass velocities is
shown as the solid
histogram (note that both distributions are normalised to unit area).
{\bf Middle panel}: the mean stellar mass as a function of
centre of mass velocity is plotted as open cicles for the initial distribution,
and as solid dots after cluster dissolution. {\bf Bottom panel}: The initial
distribution of centre of mass velocities of single stars and binaries is shown
as the thin and thick dashed histogram respectively.
The sum of these two gives the initial distribution
shown in the top panel. The distribution after
cluster disintegration of the single stars and binaries are shown as the thin
and thick solid histogram, respectively. The sum of these two gives the final
distribution shown in the top panel. The data presented here
are available in Table~A-2.

\vskip 5mm

\noindent {\bf Figure 10.} The K-band luminosity function averaged from
20~simulations (Section~6.1).
{\bf Top panel}: Counting all stars separately we obtain the luminosity
function shown as the long-dashed histogram.
Pairing all stars at random to binary systems we obtain
the initial system luminosity function shown as the dotted histogram. This
luminosity
function (neglecting pre-main sequence stellar evolution) evolves in the
environment of our $R_{0.5}=0.85$~pc dominant mode cluster to the system
luminosity function shown as the thick solid line
histogram. This luminosity function represents a mixture of single stars
(52~per cent) and
unresolved binary systems (48~per cent, assuming all bound binaries remain
unresolved),
which have the period distribution shown as the solid histogram in fig.~7 in K2
and the mass ratio distributions shown in figs.~8 and~12 in K2. It
is the luminosity function obtained from deep photographic surveys if our model
is a true representation of the Galactic field star population.
{\bf Bottom panel}: The luminosity function within the central
2~pc sphere of the dominant mode cluster after 44 initial relaxation times (see
also
Fig.~4). Counting all stars individually we obtain the long-dashed histogram,
but counting all systems we obtain the solid histogram. To show that
significant mass and system segregation and loss has occurred (see Fig.~4) we
scale the luminosity functions from the top panel to the present luminosity
functions at $M_{\rm K}\approx4$ and plot
them as curves using the same symbols as in the top panel. An observer looking
at the central
region of a dynamically highly evolved cluster would observe a luminosity
function similar to the thick solid line histogram instead of the luminosity
function shown by the solid line. The preferential loss of low-mass stars is
evident.

\vskip 5mm

\noindent {\bf Figure 11.} Comparison of our model Galactic field luminosity
functions (tabulated in table~2 in Kroupa 1995c)
with the observed Pleiades (top panel) and Hyades (bottom panel)
luminosity functions (Section~6.2.1). In both panels $\Psi_{\rm mod,sing}$
is shown as the long-dashed curve, $\Psi_{\rm mod,sys}(t=0)$
is the dotted curve and $\Psi_{\rm mod,sys}(t=1\,{\rm Gyr})$ is
shown as the solid curve.
{\bf Top panel}: The solid dots are the observed luminosity function for the
Pleiades Cluster (Hambly et al. 1991). The model luminosity
functions are plotted assuming a distance modulus $m-M=5.5$, except for the
short-long-dashed model, which is identical to the solid curve apart from
assuming
$m-M=6$. The large crosses are the photometric luminosity function for the
Galactic field (Stobie et al. 1989) transformed to $m_{\rm I}$.
{\bf Bottom panel}: The solid dots
are the observed luminosity function for the Hyades open cluster (Reid 1993).

\vskip 5mm

\noindent {\bf Figure 12.} The time evolution of the model system luminosity
function (assuming
all binaries are unresolved) in the central 5~pc sphere
of the dominant mode cluster (Section~6.2.2).
{\bf Top panel}: Assuming a distance
modulus of $m-M=6$ we plot the model luminosity functions in the I-band, and
compare with observational data of the Pleiades Cluster shown as open
and solid symbols (Hambly et al. 1991)
scaled to the models at $m_{\rm I}\approx15$. {\bf Bottom
panel}: The observed luminosity function for the Hyades Cluster (Reid 1993)
shown as open and solid symbols is scaled to the models at $M_{\rm
V}\approx12$. Taking account of the higher metallicity of the Hyades Cluster
would shift the model luminosity functions to fainter luminosities by about
0.3~mag.

\vskip 5mm

\noindent {\bf Figure 13.} Evaporation of low mass stars from the central
5~pc volume in the $R_{0.5}=0.85$~pc cluster (Section~6.2.2).
$\zeta(t)$ is defined by
equation~6. $\zeta_{\rm V}$ and $\zeta_{\rm I}$ represent our model of the
dynamical evolution of the dominant mode cluster. The Hyades
datum is the lower right cross, and the Pleiades datum is the upper left cross.

\vfill
\bye